\documentclass[12pt]{article}
\usepackage{amsfonts}
\usepackage{amsmath,amssymb}
\usepackage{times}
\usepackage{graphicx}
\usepackage{color}
\usepackage{multirow}
\usepackage{here}
\usepackage[authoryear]{natbib}
\usepackage{comment}
\usepackage[hyphens]{url}
\usepackage{hyperref}
\usepackage{rotating}
\usepackage{bbm}
\usepackage{latexsym}
  
\newcommand{\captionfonts}{\normalsize}
  
\makeatletter  
\long\def\@makecaption#1#2{%
    \vskip\abovecaptionskip
    \sbox\@tempboxa{{\captionfonts  #1:  #2}}%
    \ifdim  \wd\@tempboxa  >\hsize
	{\captionfonts  #1:  #2\par}
    \else
	\hbox  to\hsize{\hfil\box\@tempboxa\hfil}%
    \fi
    \vskip\belowcaptionskip}
\makeatother    

\begin{document}
\hspace{13.9cm}1
  
\  \vspace{20mm}\\
  
{\LARGE  Dynamical  Mechanism  of  Sampling-Based  Probabilistic  Inference  Under  Probabilistic  Population  Codes}
  
\  \\
{\bf  \large  Kohei  Ichikawa$^{\displaystyle  1,  \displaystyle  2}$}\\
{\bf  \large  Asaki  Kataoka$^{\displaystyle  1,  \displaystyle  2}$}  \\
{$^{\displaystyle  1}$Graduate  School  of  Arts  and  Sciences,  The  University  of  Tokyo.}\\
{$^{\displaystyle  2}$ACES,  Inc.}\\
%
  
  
\thispagestyle{empty}
\markboth{}{NC  instructions}
\  \vspace{-0mm}\\
%
\begin{center}  {\bf  Abstract}  \end{center}
Animals    make  efficient  probabilistic  inferences  based  on  uncertain  and  noisy  information  from  the  outside  environment.  It  is  known  that  probabilistic  population  codes,  which  have  been  proposed  as  a  neural  basis  for  encoding  probability  distributions,  allow  general  neural  networks  (NNs)  to  perform  near-optimal  point  estimation.  However,  the  mechanism  of  sampling-based  probabilistic  inference  has  not  been  clarified.  In  this  study,  we  trained  two  types  of  artificial  NNs,  feedforward  NN  (FFNN)  and  recurrent  NN  (RNN),  to  perform  sampling-based  probabilistic  inference.  Then,  we  analyzed  and  compared  their  mechanisms  of  sampling.  As  a  result,  it  was  found  that  sampling  in  RNN  was  performed  by  a  mechanism  that  efficiently  uses  the  properties  of  dynamical  systems,  unlike  FFNN.  In  addition,  it  was  found  that  sampling  in  RNNs  acted  as  an  inductive  bias,  enabling  a  more  accurate  estimation  than  in  maximum  {\it  a  posteriori}  estimation.  These  results  will  provide  important  arguments  for  discussing  the  relationship  between  dynamical  systems  and  information  processing  in  NNs.
  
\section{Introduction}
Given  ambiguous  information  from  external  environments,  humans  and  other  intelligent  species  can  perform  efficient  probabilistic  inference  \citep{KNILL2004712,ANGELAKI2009452,HAEFNER2016649,Ernst2002,Merfeld1999,Doya2007-yy},  which  may  be  achieved  by  estimating  posterior  distributions  over  the  cause  of  the  input.  One  neural  basis  for  such  an  estimation  is  {\it  probabilistic  population  codes}  (PPC),  proposed  by  \cite{Ma2006,BECK20081142,MA2008217,Orhan2017}.  In  PPC,  neural  populations  manipulate  Poisson-like  variability  to  represent  the  probability  distributions  about  the  external  environment.  Some  studies  have  suggested  that  PPC  is  used  for  inference  in  the  actual  brain\citep{Walker:2020aa,  Tanabe:2013aa}.
  
Previous  studies  have  shown  that  given  input  signals  under  PPC,  generic  neural  networks  (NNs)  can  efficiently  perform  probabilistic  inference  on  the  posterior  distribution  of  the  stimuli\citep{Orhan2017}.  Although  they  investigated  only  point  estimations,  such  as  maximum  {\it  a  posteriori}  (MAP),  as  probabilistic  inference,  some  scholars  claim  that  sampling-based  inference  in  which  outcomes  are  sampled  from  the  posterior  distribution  is  more  biologically  realistic  than  point  estimation\citep{HAEFNER2016649,  Banyai2723,  ORBAN2016530}.  As  for  sampling,  \citep{Moreno-Bote12491}  showed  that  Bayesian  sampling  from  a  categorical  distribution  can  be  achieved  by  attractor  NNs;  however,  the  case  of  continuous  distribution  was  not  investigated.
Hence,  the  mechanism  of  sampling  from  a  continuous  distribution  in  generic  NNs  and  the  relationships  to  PPC  remain  elusive.
  
Therefore,  in  this  study,  we  trained  feedforward  NNs  (FFNNs)  and  recurrent  NNs  (RNNs)  to  perform  sampling-based  inference  from  the  posterior  distribution  on  cue-combination  task\citep{cue_combination,Ernst2002},  which  is  a  common  psychophysical  experiment.  As  a  result,  we  found  that  both  NNs  could  efficiently  perform  sampling-based  probabilistic  inferences  with  PPC,  and  RNNs  are  more  accurate.  Then,  we  investigated  the  dynamical  mechanisms  of  sampling-based  inference  in  the  two  types  of  artificial  NNs  (ANNs)  and  showed  their  different  mechanisms.  Finally,  we  trained  them  for  a  point  estimation  task  and  compared  their  performance  on  it  with  sampling-based  inference.  We  found  that  they  were  more  successful  in  sampling-based  inference  than  point  estimation  and  revealed  the  reason  by  analyzing  both  geometrical  and  dynamical  features  of  their  neural  activities.    These  results  suggest  that  the  training  on  sampling-based  inference  plays  the  role  of  an  inductive  bias.  The  results  also  clarify  the  mechanism  of  sampling-based  inference  and  indicate  its  significance;  further,  they  are  essential  for  bridging  the  gap  between  dynamics  \citep{NEURIPS2019_5f5d4720,fixed_point,Vyas2020-fp}  and  geometry  \citep{PhysRevX.8.031003,  Cohen2020}  in  NNs.
  
\section{Methodology}
\subsection{Task}
We  adopt  a  cue-combination  task  in  which  two  ambiguous  cues  $c_A$  and  $c_B$  representing  information  about  the  same  environmental  status  $s$  are  integrated  to  compute  the  posterior  distribution  of  $s$.  For  instance,  this  task  can  be  considered  a  simple  task  for  a  situation  where  the  location  of  a  bird  (corresponds  to  $s$)  is  estimated  from  its  faintly  visible  sound  (corresponds  to  $c_A$)  and  audible  chirping  (corresponds  to  $c_B$)  in  a  forest.  Notably,  the  mechanism  of  solving  this  cue-combination  task  using  PPC  as  MAP  estimation  has  been  investigated\citep{Ma2006,  Orhan2017}.
  
We  assume  that  the  likelihood  of  two  cues  $p(c_A|s)$  and  $p(c_B|s)$  are  Gaussian  with  respect  to  $s$  and  their  means  and  standard  deviations  are  $(\mu_A,  \sigma_A)$  and  $(\mu_B,  \sigma_B)$,  respectively.
Assuming  that  the  prior  distribution  is  uniform,  the  posterior  calculated  by  Bayes’  rule  $p(s|c_A,  c_B)  \propto  p(c_A|s)p(c_B|s)p(s)$  also  has  a  Gaussian  distribution  $p_{\rm  post}=p(s|c_A,  c_B)  =  \mathcal{N}(\mu_{\rm  post},  \sigma^2_{\rm  post})$  with
  
\begin{align}
	\label{eq:mu_post}
	\mu_{\rm  post}  &=  \frac{\sigma_A^2}{\sigma_A^2  +  \sigma_B^2}\mu_{B}  +  \frac{\sigma_B^2}{\sigma_A^2  +  \sigma_B^2}\mu_A  \\
	\label{eq:sigma_post}
	\frac{1}{\sigma_{\rm  post}^2}  &=  \frac{1}{\sigma_A^2}  +  \frac{1}{\sigma_B^2}
\end{align}
  
Now,  we  consider  two  signals  ${\bf  x^A}$  and  ${\bf  x^B}$  encoding  $c_A$  and  $c_B$,  respectively,  by  PPC  defined  as  follows.
The  signals  ${\bf  x^A}  and  {\bf  x^B}$  are  random  variables  whose  arbitrary  element,  which  corresponds  to  a  firing  rate  of  a  single  neuron,  takes  a  value  larger  than  $0$,  and  it  follows  a  Poisson  variability  given  by  Eq.  (\ref{eq:poisson_variability1},\ref{eq:poisson_variability2}).
  
\begin{equation}
	\label{eq:poisson_variability1}
	p({\bf  x}^A|\mu_A)  =  \prod_i  \frac{e^{-f_i(\mu_A)}f_i(\mu_A)^{x^A_i}}{x^A_i!}
\end{equation}
\begin{equation}
	\label{eq:poisson_variability2}
	p({\bf  x}^B|\mu_B)  =  \prod_i  \frac{e^{-f_i(\mu_B)}f_i(\mu_B)^{x^B_i}}{x^B_i!}
\end{equation}
  
$f_i$  is  a  tuning  function  of  the  $i$-th  neuron,  which  has  a  bell-shaped  tuning  curve,  and  one  can  write  it  down  using  the  neuron's  preferred  stimulus  $\phi_i$  as  \\  $f_i(s)  =  g\exp\bigl({-(s-\phi_i)^2  /  2\sigma_{\rm  PPC}^2}\bigr)$.  preferred  stimuli  have  uniform  intervals  in  the  range  of  $\mu^{\rm  min}$–$\mu^{\rm  max}$  as  $\phi_i  =  \mu^{\rm  min}  +  i/m\times  (\mu^{\rm  max}  -  \mu^{\rm  min})  $.
Gain  $g$  represents  the  reliability  of  the  input  signal\citep{TOLHURST1983775}.
The  gain  is  related  to  the  variability  $\sigma^2$  of  the  input  signal  as  follows:  $g  =  1/\sigma^2$.
One  trial  lasts  for  $80  \tau$,  where  $\tau$  is  a  unit  time.
See  Appendix  A  for  specific  examples  of  the  input  signals.
  
During  trials,  the  concatenation  of  signals  ${\bf  x^A}=(x^A_1,x^A_2,...,x^A_m)$  and  \\  ${\bf  x^B}=(x^B_1,x^B_2,...,x^B_m)$  as  ${\bf  x}  =  (x^A_1,...,x^A_m,  x^B_1,...,x^B_m)$  is  inputted  to  an  NN.  In  this  task,  the  NN  has  to  estimate  the  statistics  of  two  cues,  compute  Eqs.  (\ref{eq:mu_post})  and  (\ref{eq:sigma_post}),  and  then  generate  output  signals  as  sampling  from  the  posterior  distribution.
  
\subsection{Neural  Networks(NNs)}
\begin{figure}
	\centering
	\includegraphics[width=13.5cm]{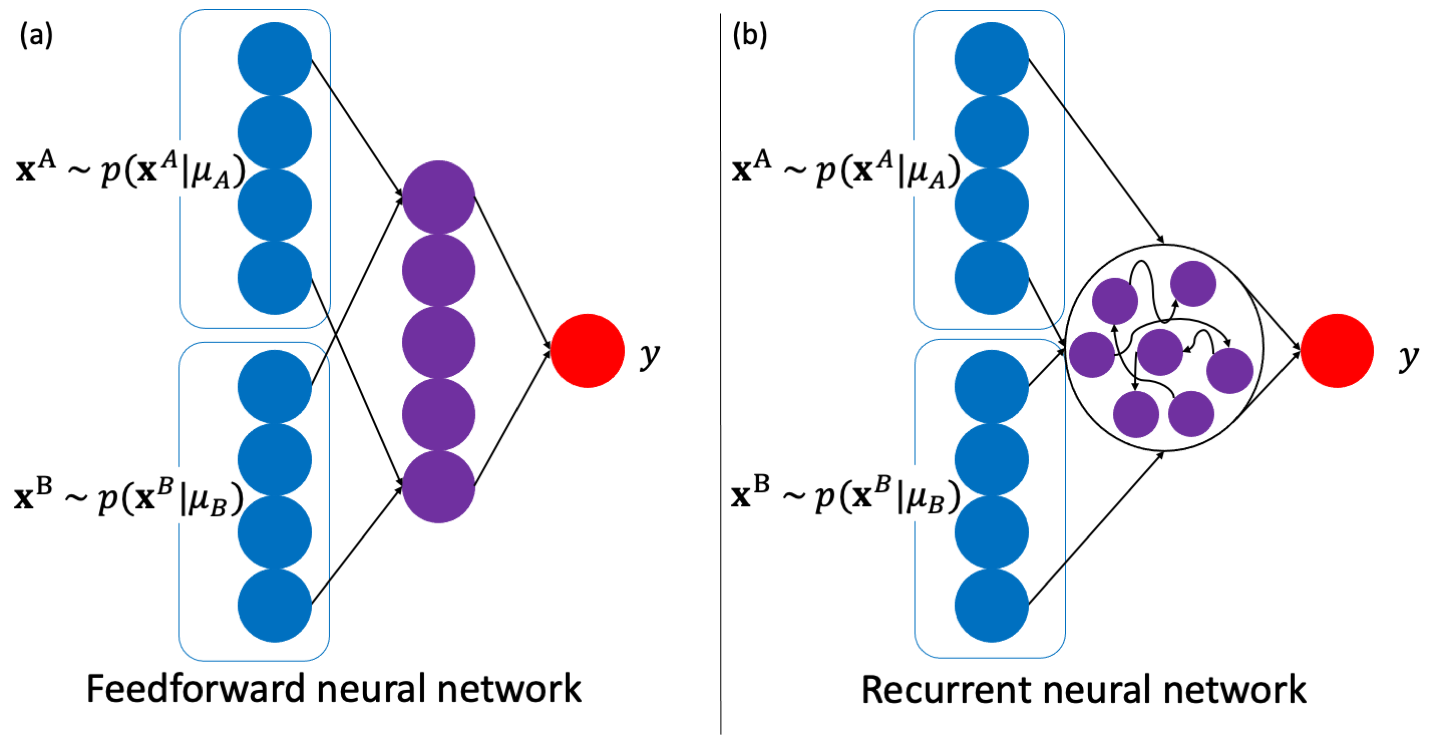}
	\caption{Network  architectures.  (a)  FFNN  has  three  layers.  Input  layers  encode  $c_A$  and  $c_B$  by  probabilistic  population  codes.  (b)  Similar  to  FFNN,  RNN  comprises  three  layers,  but  the  hidden  layer  has  internal  connections  and  constitutes  a  dynamical  system.}
	\label{fig:fig_networks}
\end{figure}

:  
We  trained  two  different  types  of  ANNs  on  the  cue-combination  task  above:  FFNN  \citep{SVOZIL199743}  and  RNN  \citep{BARAK20171}.  See  Fig.  \ref{fig:fig_networks}  for  schematic  illustrations  of  the  ANNs.
  
\subsubsection{FFNN}
We  employed  a  3-layer  FFNN  comprising  input,  hidden,  and  output  layers.
The  nonlinear  activation  function  in  the  hidden  layer  is  rectified  linear  unit  (ReLU)  \citep{ReLU}.
Given  an  input  signal  $\bf  x$,  the  hidden  and  output  layer  activations,  ${\bf  h}(t)$  and  $y(t)$,  respectively,  are  deterministically  computed  as  follows:
  
\begin{align}
	{\bf  h}(t)  &=  {\rm  ReLU}(W_{\rm  in}{\bf  x}(t)  +  {\bf  b}_{\rm  in}),  \\
	\label{eq:output_computation}
	y(t)  &=  {\rm  clipping}[W_{\rm  out}({\bf  h}(t)  +  {\bf  b}_{\rm  out})]
\end{align}
where
\begin{equation}
	\label{eq:clipping}
	{\rm  clipping}(z)  =  \left\{  \begin{array}{ll}
        	-20  &  (z  <  -20)  \\
        	z  &  (-20  \leq  z  \leq  20)  \\
        	20  &  (z  >  20)
	\end{array}
	\right.
\end{equation}
  
\subsubsection{RNN}
The  emerging  dynamics  of  ${\bf  h}(t)$  in  the  hidden  layer  of  RNN  is  defined  as  follows:
  
\begin{equation}
	{\bf  h}(t+1)  =  (1-\alpha){\bf  h}(t)+\alpha  {\rm  ReLU}(W_{\rm  rec}{\bf  h}(t)+W_{\rm  in}{\bf  x}(t)+{\bf  b})
\end{equation}

where  we  set  $\alpha  =  0.25$.
The  output  layer  activation  is  the  same  as  Eq.  (\ref{eq:output_computation}).

\subsection{Training}
  
As  a  loss  function  ${\mathcal  L}$  in  the  training  phase,  we  use  the  Kullback–Leibler  divergence  (KLD)  \citep{Qamar20332}  defined  for  two  probability  distributions  $p(s|c_A,c_B)$  and  $q_{\rm  dis}$  sampled  from  the  output  $y(t)$.  For  simplicity,  we  discretized  the  range  of  possible  values  of  $y$,  $-20\leq  y  <20$,  into  200  bins  and  calculated  the  KLD  for  the  discrete  probability  distribution.
\begin{equation}
	\label{eq:loss}
	{\mathcal  L}  =  \sum_k^{200}  q_{\rm  dis}(k)\log\frac{q_{\rm  dis}(k)}{p_{\rm  post}(k)}
\end{equation}
  
where  $q_{\rm  dis}(k)$  is  defined  as  follows:
\begin{equation}
	q_{\rm  dis}(k)  =  \frac{1}{T}\sum_{t=1}^{T}  {\rm  I}_k[y(t)]
\end{equation}
  
${\rm  I}_k[y(t)]$  is  an  indicator  that  $y(t)$  is  in  the  $k$-th  bin.
\begin{equation}
	\label{eq:indicator}
	{\rm  I}_k[y(t)]  =  \left\{  \begin{array}{ll}
        	1  &  \  (\frac{k-1}{200}\times  40-20  \leq  y(t)  <  \frac{k}{200}\times  40-20)  \\
        	0  &  \    (\rm  otherwise)
        	\end{array}
        	\right.
\end{equation}
  
As  we  intend  to  train  NNs  using  the  gradient  descent  scheme  \citep{DeepLearning},  the  loss  function  should  be  differentiable.  However,  ${\mathcal  L}$  is  not  differentiated  because  of  Eq.  (\ref{eq:indicator}).  Therefore,  we  defined  a  “soft”  distribution  $q_{\rm  soft}$,  which  is  differentiable  and  an  approximation  of  $q_{\rm  dis}$:
  
\begin{equation}
	q_{\rm  soft}(k)  =  \frac{1}{T}\sum_{t=k}^{T}\frac{-\tanh[\beta((y(t)-a_k)^2-1/4)]+1}{2}
\end{equation}
  
where  $a_k  =  \frac{2k-1}{400}\times  40-20  $,  and  $\beta=2$.
Then,  we  calculate  KLD  for  $q_{\rm  soft}$  and  $p_{\rm  post}$  as  a  loss  function.
\begin{equation}
	\label{eq:des_loss}
	L  =  \sum_k^{200}  q_{\rm  soft}(k)\log\frac{q_{\rm  soft}(k)}{p_{\rm  post}(k)}
\end{equation}
  
We  trained  the  NNs  using  Adam  \citep{adam}  as  a  specific  optimization  algorithm.
The  batch  size  and  value  of  weight  decay  were  set  to  50  and  0.001,  respectively,  and  the  training  was  performed  for  6,000  iterations.
The  source  codes  of  the  simulations  in  this  article  are  available  at  \hyperref[url]{https://github.com/tripdancer0916/probabilistic-inference-rnn}.
  
\begin{table}[H]
	\caption{Hyperparameters}
	\begin{center}
        	\begin{tabular}{c|c}  \hline\hline
                	Attribute  &  Value  \\  \hline\hline
	                Range  of  $\mu_A,\mu_B$  &  $-20\leq  \mu_A,\mu_B\leq  20$    \\
                	Range  of  $g_A,g_B$  &  $0.25  \leq  g_A,g_B\leq  1.25$  \\
                	Length  of  ${\bf  x}(t)$  &  $2m=200$  \\
                	$\sigma_{\rm  PPC}^2$  &  5  \\
                	Lasting  time  of  ${\bf  x}(t)$  &  $T=80$  \\
                	\#Neurons  in  the  hidden  layer  &  300  \\
                	$\alpha$  &  0.25  \\
                	Batch  size  &  50  \\
                	Optimization  algorithm  &  Adam  \\
                	Learning  rate  &  0.001  \\
                	Iteration  &  6000  \\
                	Weight  decay  &  0.0001  \\  \hline
                	\end{tabular}
	\end{center}
\end{table}
  
\section{Results}
\begin{figure}[H]
	\centering
	\includegraphics[width=13.5cm]{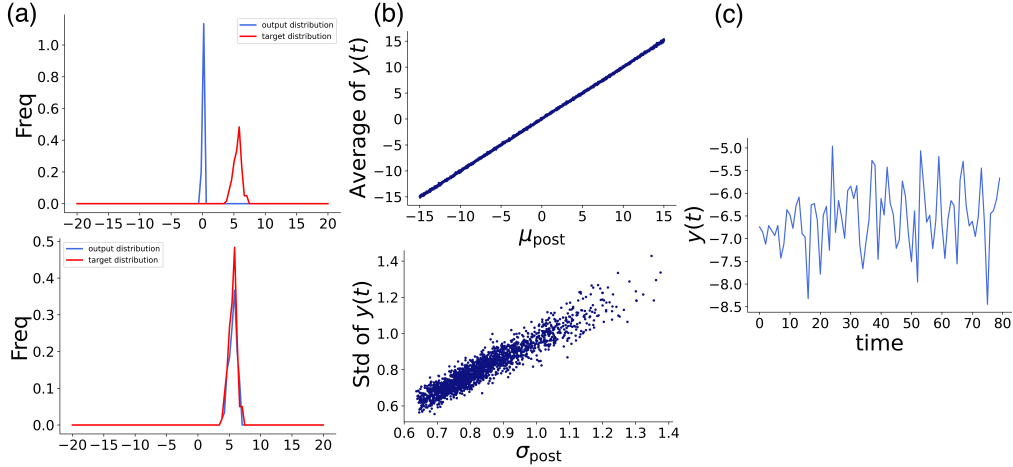}
	\caption{The  behavior  of  trained  RNN  with  cue-combination  task.  (a)  Comparison  between  the  posterior  distribution  and  histogram  of  the  output  $y(t)$  of  the  RNN  after  training.  (b)  (top)  Comparison  between  the  means  of  the  posterior  distribution  $\mu_{\rm  post}$  and  output  $y(t)$.  (bottom)  Comparison  between  the  standard  deviations  of  the  posterior  distribution  $\sigma_{\rm  post}$  and  output  $y(t)$.  Each  point  corresponds  to  several  input  signals  ${\bf  x}$.  Here,  $\mu_A$  and  $\mu_B$  satisfies  $-15\leq  \mu_A,\mu_B  \leq  15$.  (c)  The  dynamics  of  the  output  $y(t)$.  One  can  see  that  a  stochastic  output  appears,  neither  persistent  nor  periodic.}
	\label{fig:training_results}
\end{figure}

As  a  result  of  training,  the  distribution  of  the  network  output  $y(t)$  for  both  FFNN  and  RNN  approached  the  posterior  distribution  $p_{\rm  post}$  [Fig.  \ref{fig:training_results}(a)].
The  mean  and  variance  of  $y(t)$  were  similar  to  those  obtained  from  Eqs.  (\ref{eq:mu_post},  \ref{eq:sigma_post})  (Fig.  \ref{fig:training_results}(b)).
In  addition,  Fig.  \ref{fig:training_results}(c)  shows  the  output  timeseries  is  stochastically  generated,  not  periodically.
These  results  suggest  that  the  ANNs  successfully  learned  to  perform  sampling-based  inference  from  the  posterior  distribution  in  the  cue-combination  task.  We  also  confirmed  that  these  ANNs  could  learn  to  perform  sampling-based  probabilistic  inference  in  another  cognitive  task  (see  Appendix  B).
  
\begin{figure}
	\centering
	\includegraphics[width=13.5cm]{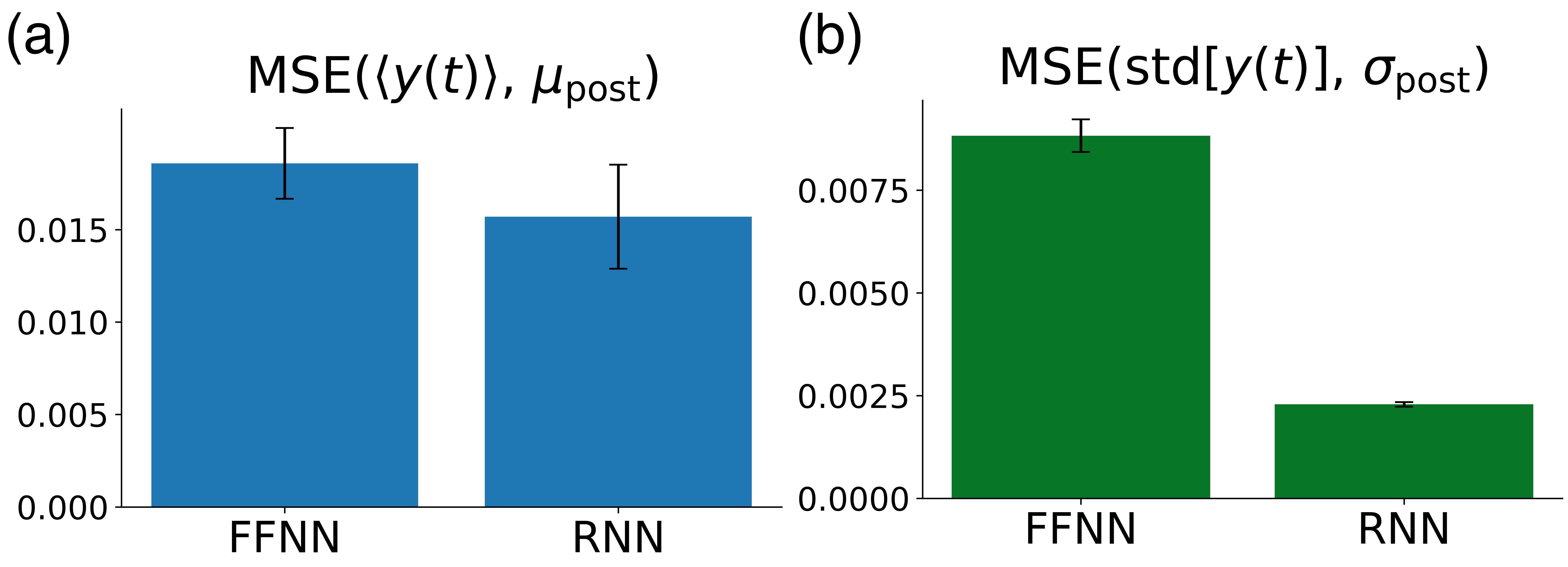}
	\caption{Comparison  of  accuracy  between  the  trained  RNN  and  FFNN.  (a)  Mean  squared  error  (MSE)  of  the  means  of  the  posterior  distribution  $\mu_{\rm  post}$  and  output  $\langle  y(t)  \rangle$($p=0.0879$).  (b)  MSE  of  the  standard  deviations  of  the  posterior  distribution  $\sigma_{\rm  post}$  and  output  ${\rm  std}[y(t)]$($p=5.87\times  10^{-5}$).  Though  there  is  no  significant  difference  between  FFNN  and  RNN  for  $\langle  y(t)  \rangle$,  RNN  is  significantly  more  accurate  for  ${\rm  std}[y(t)]$.  The  bars  are  the  average  MSE  of  three  models  trained  under  the  same  condition,  and  the  error  bars  represent  the  standard  deviation.}
	\label{fig:training_results_mean_std}
\end{figure}

Here,  accuracy  computed  by  comparing  the  mean  and  the  variance  of  the  outputs  of  Eqs.  (\ref{eq:mu_post})  and  (\ref{eq:sigma_post})  in  terms  of  mean  squared  error  (MSE)  were  higher  in  the  RNN  than  in  the  FFNN,  especially  in  the  variance,  suggesting  that  the  RNN  performed  sampling-based  inference  more  efficiently  (Fig.\ref{fig:training_results_mean_std}).
We  also  found  that  the  FFNN  and  RNN  could  perform  efficient  information  processing  from  the  perspective  of  information  theory  and  confirm  that  RNN  can  more  efficiently  process  input  information  than  FFNN(See  Appendix  C).
  
In  this  section,  the  analyses  of  the  mechanism  of  sampling  behavior  in  the  FFNN  and  RNN  are  described.

\subsection{Sampling  mechanism  in  the  FFNN}
\begin{figure}
	\centering
	\includegraphics[width=13.5cm]{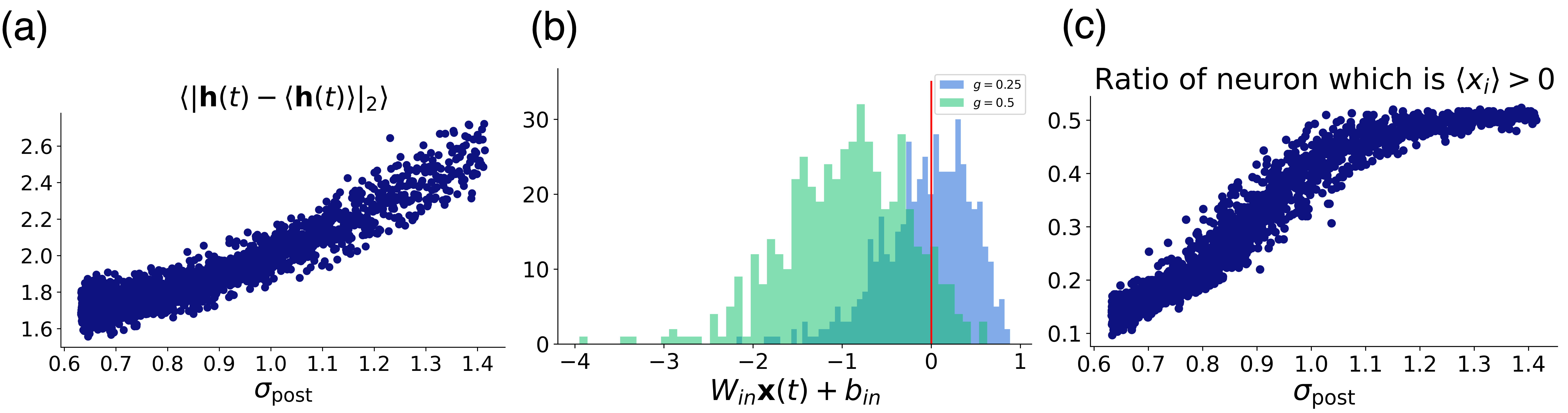}
	\caption{Sampling  mechanism  in  FFNN.  (a)  Relation  between  the  standard  deviation  $\sigma_{\rm  post}$  of  the  posterior  distribution  and  the  fluctuation  in  hidden  layer  activation.  The  fluctuation  is  defined  as  time-averaged  distance  $\langle  |{\bf  h}(t)  -  \langle  {\bf  h}(t)  \rangle|_2  \rangle$  of  the  activation  from  the  mean  activity  $\langle  {\bf  h}(t)  \rangle$.
	There  is  a  positive  correlation  between  the  variance  of  the  posterior  distribution  and  the  fluctuation  of  the  hidden  layer  activities.
	(b)  Frequency  distribution  of  element-wise  values  of  linear  mapping  from  the  input  signal  into  the  hidden  layer.  As  $g$  increases,  the  entire  distribution  shifts  toward  a  negative  direction.  Notably,  the  input  signals  lasted  for  $500  \tau$  in  this  visualization.
	(c)  Rate  of  neurons  with  values  greater  than  0  to  the  entire  number  of  hidden  layer  neurons.  It  increases  with  $\sigma_{\rm  post}$  (and  equivalently  $g_{A,B}$  decreases).}
	\label{fig:ffnn_sampling_mechanism}
\end{figure}

To  accurately  perform  sampling  from  the  posterior  distribution,  the  NN  needs  to  manipulate  the  output  $y(t)$  variability  using  stochastic  perturbations  of  the  input  signals.
Here,  as  shown  in  Eqs.  (\ref{eq:poisson_variability1},  \ref{eq:poisson_variability2}),  the  input  signals  have  Poisson-like  variability,  and  their  mean  and  variance  increase  linearly  with  the  gain  $g_A,  g_B$.
However,  as  the  gain  $g_A,  g_B$  corresponds  to  the  input  signals’  reliability,  the  variance  of  the  posterior  distribution  decreases  as  $g_A,  g_B$  increases.
It  means  that  the  network  should  have  an  internal  mechanism  to  reverse  the  magnitude  of  variances  of  the  input  and  output.
Here,  we  clarify  the  mechanism  that  transforms  the  input  signal  variance  to  output  variance  in  FFNN.
First,  there  are  two  possible  candidates  for  where  the  variance  transformation  occurs:  the  input  layer  to  the  hidden  layer  and  the  hidden  layer  to  the  output  layer.
As  the  magnitude  of  the  fluctuation  of  the  hidden  layer  activity  ${\bf  h}(t)  =  {\rm  ReLU}(W_{\rm  in}{\bf  x}(t)+{\bf  b}_{in})$  strictly  correlates  to  the  standard  deviation  of  the  posterior  distribution  [Fig.  \ref{fig:ffnn_sampling_mechanism}(a)],  this  variance  transformation  is  supposed  to  occur  when  the  signal  is  transformed  from  the  input  layer  to  the  hidden  layer.
In  the  following,  we  reveal  the  mechanism  of  this  variance  transformation  by  focusing  on  the  transformation  of  signals  ${\bf  x}(t)$  from  the  input  layer  to  the  hidden  layer.
  
Focusing  on  a  single  neuron  in  the  hidden  layer,  we  investigated  how  differences  in  the  gain  of  the  input  signal  affect  the  neuronal  activity  in  that  hidden  layer.  For  simplicity,  we  set  $g_1=g_2=g$.  Comparing  the  results  for  $g=0.25$  and  $g=0.5$,  we  found  that  the  distribution  of  the  amount  of  activity  obtained  by  the  linear  transformation  $(W_{in}{\bf  x}(t)+{\bf  b}_{in})$  from  the  input  layer  widened  and  shifted  to  the  left  as  $g$  increased  [Fig.  \ref{fig:ffnn_sampling_mechanism}(b)].  Indeed,  this  tendency  is  supported  by  the  statistics  of  the  input  matrix  $W_{\rm  in}$  and  the  bias  ${\bf  b}_{\rm  in}$;  the  mean  values  of  $W_{\rm  in}$  and  ${\bf  b}_{\rm  in}$,  respectively,  satisfy  $E[W_{\rm  in}]  <  0$  and  $E[{\bf  b}_{\rm  in}]  >  0$.
Here,  considering  $\rm  ReLU$  nonlinearity,  the  variance  of  the  hidden  layer  is  affected  only  by  hidden  neurons  with  activities  greater  than  $0$,  from  which  a  negative  correlation  between  $g$  and  ${\rm  ReLU}(W_{in}{\bf  x}(t)+{\bf  b}_{in})$  can  be  derived.

Assuming  that  the  above  mechanism  is  in  the  FFNN,  the  number  of  hidden  neurons  with  values  greater  than  $0$  decreases,  i.e.,  the  hidden  layer  gets  sparser.
In  fact,  an  analysis  shows  that  a  sparsity-based  coding  is  present  [Fig.  \ref{fig:ffnn_sampling_mechanism}(c)].  This  sparsity-based  coding  is  similar  to  the  coding  in  MAP  estimation  by  FFNN  \citep{Orhan2017}.
  
\subsection{Sampling  mechanism  in  the  RNN}
\begin{figure}
	\centering
	\includegraphics[width=14.5cm]{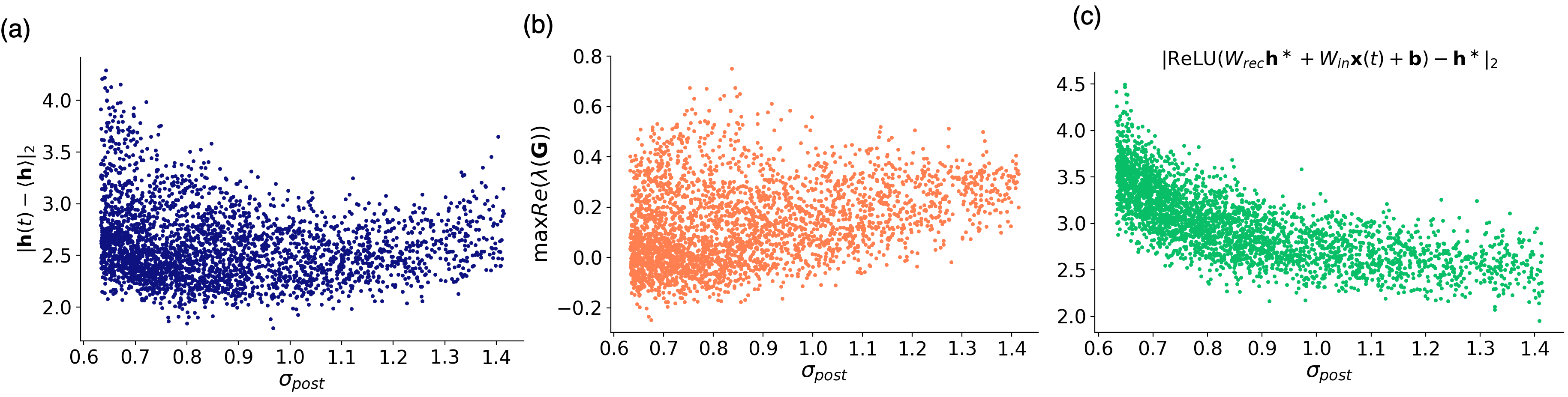}
	\caption{Sampling  mechanism  in  RNN.  (a)  Relation  between  the  standard  deviation  of  the  posterior  distribution  $\sigma_{\rm  post}$  and  the  fluctuation  of  the  hidden  layer  activities.  Different  from  the  case  of  the  FFNN,  the  hidden  layer  activities  nonmonotonically  change  with  the  posterior  deviation.  (b)  The  real  parts  of  eigenvalues  of  the  Jacobian  matrix  near  fixed  points.  They  positively  correlate  with  the  fluctuation  of  the  hidden  layer  activities  when  $\sigma_{\rm  post}$  is  large.  (c)  Fluctuations  of  the  input  signal.  They  positively  correlate  with  the  fluctuation  of  the  hidden  layer  activities  when  $\sigma_{\rm  post}$  is  small.}
	\label{fig:rnn_sampling_mechanism}
\end{figure}

Now,  we  discuss  the  sampling  mechanism  in  the  RNN.
In  the  RNN,  the  relationship  between  the  fluctuation  of  the  hidden  layer  firing  rate  ${\bf  h}(t)$  and  the  standard  deviation  $\sigma_{\rm  post}$  of  the  posterior  distribution  is  nonmonotonic,  which  differs  from  the  result  in  the  FFNN  [(Fig.\ref{fig:rnn_sampling_mechanism}(a)].
Here,  the  variation  of  ${\bf  h}(t)$  and  the  standard  deviation  $\sigma_{\rm  post}$  are  negatively  correlated  in  the  region  where  $\sigma_{\rm  post}$  is  small,  whereas  they  are  positively  correlated  in  the  region  where  $\sigma_{\rm  post}$  is  large.
  
First,  we  found  that  when  the  input  signals  were  stationary,
the  dynamics  of  the  hidden  layer  converged  to  certain  fixed  points.
Each  of  these  fixed  points  corresponds  to  a  pair  of  mean  and  standard  deviation  of  the  posterior  distribution  $(\mu_{\rm  post},  \sigma_{\rm  post})$.
Hence,  we  can  decompose  the  dynamics  of  the  hidden  layer  to  fixed-point  neural  states  and  the  fluctuation  around  them.
Assuming  that  the  fluctuation  is  sufficiently  small,
we  can  linearize  the  dynamics  around  a  fixed  point  \citep{fixed_point},  and  it  yields
\begin{equation}
	d{\bf  h}=\left.\frac{d{\bf  F}}{d{\bf  h}}\right|_{\bf  h=h^\ast}({\bf  h}-{\bf  h}^\ast)dt+dW
\end{equation}
where
\begin{equation*}
	F({\bf  h})  \equiv  {\rm  ReLU}(W_{\rm  rec}{\bf  h}  +  W_{\rm  in}\overline{\bf  x}  +  {\bf  b}_{\rm  in})
\end{equation*}
and
\begin{equation*}dW  \simeq  {\rm  ReLU}(W_{rec}{\bf  h}^\ast  +W_{\rm  in}  {\bf  x}(t)+{\bf  b})  -  {\rm  ReLU}(W_{rec}{\bf  h}^\ast  +W_{\rm  in}  {\bf  \bar  x}+{\bf  b})\equiv  d\tilde  W.
\end{equation*}
  
The  term  $d{\bf  h}=\left.\frac{d{\bf  F}}{d{\bf  h}}\right|_{\bf  h=h^\ast}$  represents  a  Jacobian  matrix  near  the  corresponding  fixed  point  ${\bf  h}^\ast$,  which  determines  the  characteristics  of  the  landscape  of  neural  dynamics  \citep{NEURIPS2019_5f5d4720}.
Meanwhile,  the  term  $dW$  corresponds  to  the  fluctuation  of  the  input  signal.
It  is  conceivable  that  the  mechanism  underlying  sampling-based  inference  in  the  RNN  is  a  combination  of  effects  from  these  two  terms.
  
Hence,  we  investigated  the  relationships  among  the  largest  eigenvalues  of  the  Jacobian  matrix  ${\bf  G}  =  \left.\frac{d{\bf  F}}{d{\bf  h}}\right|_{\bf  h=h^\ast}$  near  each  fixed  point,
the  fluctuation  $d\tilde  W$  of  the  hidden  layer  activities,
and  the  standard  deviation  $\sigma_{\rm  post}$  [Fig.  \ref{fig:rnn_sampling_mechanism}(b,  c)].
  
Comparing  this  result  with  the  relationship  between  the  fluctuation  of  the  hidden  layer  firing  rate  ${\bf  h}(t)$  and  standard  deviation  $\sigma_{\rm  post}$  [Fig.  \ref{fig:rnn_sampling_mechanism}(a)],  in  the  region  where  $\sigma_{\rm  post}$  is  small,  the  trend  is  similar  to  the  change  in  $d\tilde  W$  [Fig.  \ref{fig:rnn_sampling_mechanism}(b)],  whereas,  in  the  region  where  $\sigma_{\rm  post}$  is  large,  the  trend  is  similar  to  the  change  in  the  largest  eigenvalues  of  the  Jacobian  matrix  ${\bf  G}$  [Fig.  \ref{fig:rnn_sampling_mechanism}(c)].
  
In  the  following,  we  discuss  the  mechanism  of  controlling  the  variance  of  the  output  $y(t)$  in  each  of  the  two  areas:  small-  and  large-variance  areas.

\subsubsection{Small-variance  area}
When  the  standard  deviation  $\sigma_{\rm  post}$  of  the  posterior  distribution  is  small,  different  from  the  FFNN  case,  the  posterior  deviation  and  the  fluctuation  of  ${\bf  h}(t)$  in  RNN  positively  correlate.
It  may  naively  seem  that  the  variance  is  incontrollable.  However,  as  the  output  signal  is  determined  as  an  inner  product  of  the  hidden  activity  ${\bf  h}(t)$  and  readout  vector  ${\bf  W}_{\rm  out}$,  the  direction  of  the  fluctuation  of  ${\bf  h}(t)$  also  needs  to  be  considered,  and  it  can  explain  the  mechanism.
Let  us  consider  a  matrix  ${\bf  H}  \in  \mathbb{R}^{N\times  T}$  given  by  $T$  temporally  ordered  hidden  activities  ${\bf  h}(t)$  and  perform  eigenvalue  decomposition  of  a  covariance  matrix  ${\bf  C}\equiv  \frac{1}{T}{\bf  HH}^\mathsf{T}  -  \langle{\bf  h}\rangle\langle{\bf  h}\rangle^\mathsf{T}$  as  follows:
  
\begin{equation}
	{\bf  C}  =  \frac{1}{T}\sum_{i=1}^{N}  R_i^2  {\bf  v}_i  {\bf  v}_i^\mathsf{T}
\end{equation}
where  $R_1^2\geq  R_2^2  \geq  \cdots$.
Each  eigenvector  ${\bf  v}_i$  corresponds  to  a  principal  component  (PC)  of  $\bf  H$  \citep{PCA}.
  
\begin{figure}
	\centering
	\includegraphics[width=11.5cm]{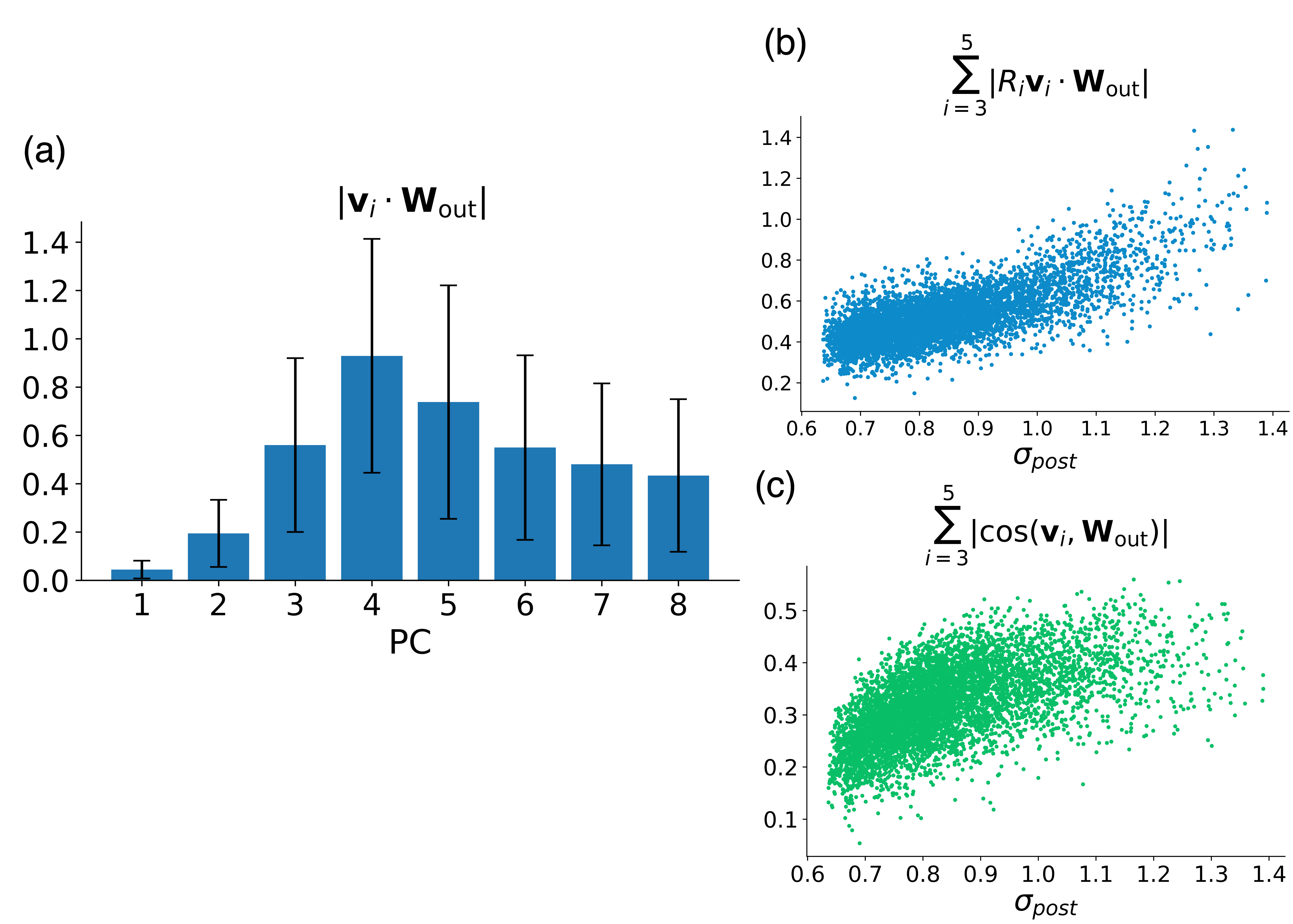}
	\caption{(a)  Absolute  inner  products  of  ${\bf  W}_{\rm  out}$  and  eigenvectors  of  the  covariance  matrix  of  the  internal  dynamics  ${\bf  h}(t)$.  The  eigenvectors  are  standardized  with  respect  to  effects  on  output  fluctuation.  Means  and  standard  deviations  from  3,000  samples  are  plotted.  (b)  The  sum  of  inner  products  of  ${\bf  W}_{\rm  out}$  and  the  third  to  fifth  eigenvectors.  They  strongly  correlate  with  $\sigma_{\rm  post}$.  (c)  The  sum  of  cosine  similarities  between  ${\bf  W}_{\rm  out}$  and  the  third  to  fifth  eigenvectors.  Their  correlation  with  posterior  variance  decreases  as  $\sigma_{\rm  post}$  increases.}
	\label{fig:small_sigma_post}
\end{figure}

We  elucidate  how  each  PC  affects  the  variance  of  the  output  $y(t)$.
As  shown  in  Fig.  \ref{fig:small_sigma_post}(a),  the  first  and  second  components  are  nearly  orthogonal  to  ${\bf  W}_{\rm  out}$,  which  implies  they  slightly  affect  the  variance  of  the  output.
Meanwhile,  the  third  to  fifth  PCs  significantly  affect  the  output  variance  via  solid  correlation  from  the  sum  of  inner  products  with  ${\bf  W}_{\rm  out}$  to  $\sigma_{\rm  post}$  (fig.  \ref{fig:small_sigma_post}(b)).
Further,  quantifying  the  fluctuation  directions  of  the  internal  dynamics  ${\bf  h}(t)$  as  angles  between  ${\bf  W}_{\rm  out}$  and  eigenvectors  of  these  (from  the  third  to  fifth)  PCs  shows  that  these  directions  negatively  correlate  with  the  posterior  variance  $\sigma_{\rm  post}$  in  the  area  where  $\sigma_{\rm  post}$  is  small.
Although  the  effect  on  the  output  $y(t)$  is  determined  by  the  magnitude  of  the  fluctuations  of  ${\bf  h}(t)$  and  their  directions,  the  results  show  that  when  $\sigma_{\rm  post}$  is  relatively  small,  the  output  variance  is  controlled  by  the  directions  of  the  fluctuations  [Fig.  \ref{fig:small_sigma_post}(c)].

\subsubsection{Large-variance  area}
In  the  area  with  large  posterior  variance,
the  correlation  between  the  variance  $\sigma_{\rm  post}$  and  angles  of  PCs  and  the  readout  vector  ${\bf  W}_{\rm  out}$  is  weak,  which  implies  that,  in  the  large-variance  area,
the  output  variance  is  simply  controlled  by  the  magnitude  of  the  fluctuation  of  the  internal  dynamics  ${\bf  h}(t)$.
In  fact,  Fig.  \ref{fig:rnn_sampling_mechanism}(a)  shows  a  positive  correlation  between  $\sigma_{\rm  post}$  and  fluctuation  of  ${\bf  h}(t)$  in  the  large-variance  area.
However,  the  input  variance  does  not  correlate  with  $\sigma_{\rm  post}$,  which  means  that  the  landscape  of  the  internal  dynamics  controls  the  output  fluctuation.
The  largest  eigenvalue  of  the  Jacobian  matrix  ${\bf  G}  =  \left.  \frac{d{\bf  F}}{d{\bf  h}}  \right|_{\bf  h=h^\ast}$  represents  stability  against  noise  around  fixed  point  ${\bf  h}^\ast$;
under  a  constant  amplitude  of  noise,  the  larger  this  value  is,  the  larger  the  fluctuation  of  the  internal  dynamics.
Fig.  \ref{fig:rnn_sampling_mechanism}  shows  the  correlation  between  the  largest  eigenvalues  and  posterior  variance  $\sigma_{\rm  post}$,  suggesting  that  they  control  the  fluctuations  of  ${\bf  h}(t)$.
  
\vspace{1.5cm}
  
\begin{figure}
	\centering
	\includegraphics[width=11.5cm]{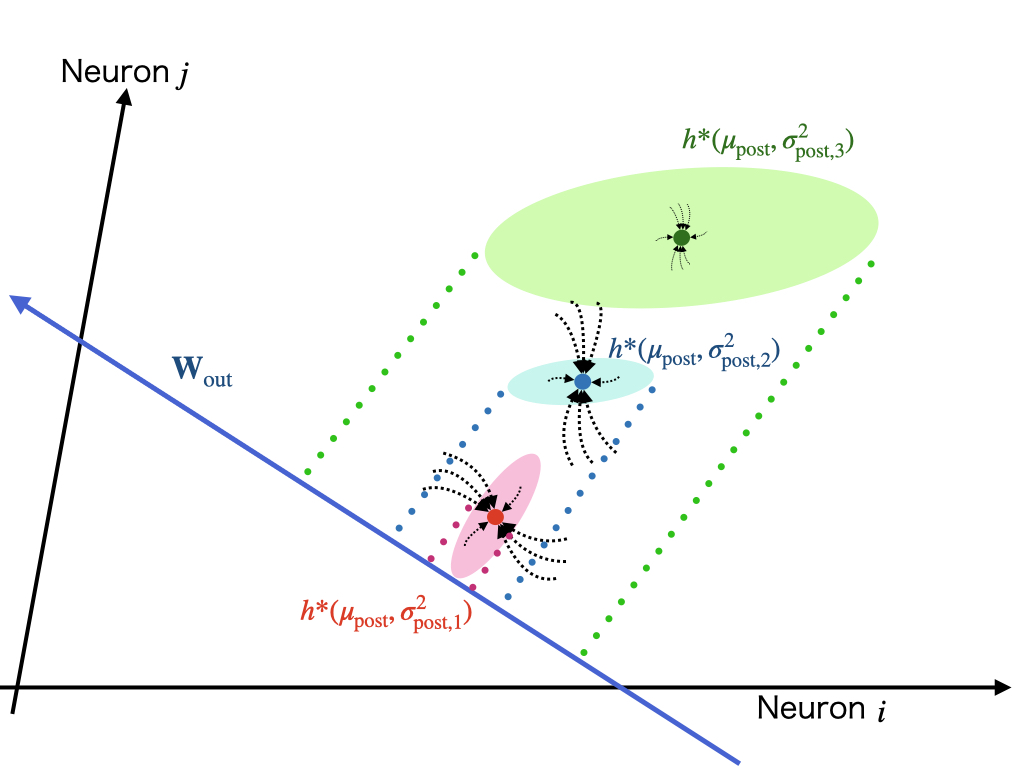}
	\caption{A  schematic  illustration  of  the  dynamical  mechanism  of  controlling  the  sampling  deviation.  The  colored  dots  are  fixed  points  corresponding  to  an  equal  mean  of  the  posterior  distribution  $\mu_{\rm  post}$  and  different  variances.  The  black  dashed  curves  represent  typical  trajectories  or  the  stability  of  the  fixed  points.  When  the  variance  of  the  posterior  distribution  is  small,  a  transition  from  a  smaller  ($\sigma_{{\rm  post},  1}$)  to  a  larger  ($\sigma_{{\rm  post},  2}$)  is  controlled  by  the  angle  of  the  readout  vector  and  the  direction  in  which  the  state  around  each  fixed  point  perturbs.  As  a  result  of  rotation,  the  variance  of  the  projection  of  the  neural  state  onto  the  readout  vector  is  enlarged.  Meanwhile,  in  the  large-variance  area,  the  transition  from  $\sigma_{{\rm  post},  2}$  to  $\sigma_{{\rm  post},  3}$  is  controlled  by  the  landscape  around  the  fixed  points.  As  the  posterior  variance  increases,  the  stability  of  the  corresponding  fixed  points  weakens,  thereby  decreasing  the  variance  of  the  projection  of  the  state  onto  the  readout  vector.}
	\label{fig:schema_sampling_dynamics}
\end{figure}

Summarizing  the  above  results,
two  attributes  contribute  to  the  control  of  sampling  in  RNNs  (Fig.\ref{fig:schema_sampling_dynamics}):
  
\begin{itemize}
\item  Small-variance  area:  The  magnitude  of  the  fluctuations  of  the  internal  dynamics  is  predominantly  determined  by  the  fluctuations  of  the  input  signal,  and  the  variance  of  the  output  signal  is  controlled  by  the  “direction”  of  the  fluctuations  of  the  internal  dynamics.
\item  Large-variance  area:  The  magnitude  of  the  fluctuations  of  the  internal  dynamics  is  predominantly  determined  by  the  landscape  of  dynamics  around  the  fixed  point,  and  the  variance  of  the  output  signal  is  controlled  by  the  “magnitude”  of  the  fluctuations  of  the  internal  dynamics.
\end{itemize}
  
RNNs  can  reasonably  perform  more  accurate  inferences  than  FFNNs  in  which  only  input  fluctuation  contributes  to  controlling  sampling.
  
Since  RNNs  have  a  larger  number  of  parameters  than  FFNNs  when  the  numbers  of  neurons  are  fixed,
we  also  tested  their  accuracy  under  a  fixed  number  of  parameters.
The  results  still  indicated  the  superiority  of  the  RNN,  implying  that  the  superiority  is  not  simply  caused  by  the  richness  of  parameters  (see  Appendix  D).
  
\subsection{Sampling  as  inductive  bias}
This  subsection  shows  the  results  of  a  comparison  of  sampling-based  probabilistic  inference  with  point  estimation.
Here,  we  used  MAP  estimation  as  an  example  of  point  estimation  and  trained  NNs  to  output  $\mu_{\rm  post}$.
Notably,  Eq.  (\ref{eq:mu_post})  claims  MAP  estimation  still  requires  NNs  to  consider  standard  deviations  $\sigma_A,  \sigma_B$.
The  loss  function  $L$  for  MAP  estimation  is  defined  as  mean  squared  error(MSE):
  
\begin{equation}
	L  =  \frac{1}{T}\sum_t  (y(t)-  \mu_{\rm  post})^2
\end{equation}
  
\begin{figure}[H]
	\centering
	\includegraphics[width=14.5cm]{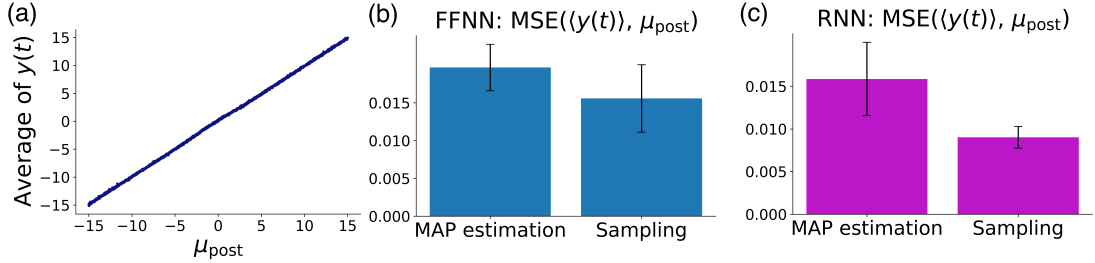}
	\caption{(a)  Comparison  between  the  mean  value  of  the  posterior  distribution  and  the  output  of  the  networks.  (b,  c)  Mean  squared  error  between  the  posterior  mean  $\mu_{\rm  post}$  and  mean  $\langle  y(t)  \rangle$  of  the  outputs.  (b)  The  results  of  FFNN.  There  is  no  significant  difference  between  MAP  inference  and  sampling-based  inference  ($p=0.421$).  (c)  The  results  of  RNN.  In  RNN,  sampling-based  inference  is  significantly  more  accurate  than  MAP  inference  ($p=0.0247$).}
	\label{fig:training_results_point}
\end{figure}

This  setting  has  already  been  studied  and  shown  to  be  possible  to  learn  for  ANNs  \citep{Orhan2017}.
In  this  study,  we  also  confirmed  the  performance  of  NNs  on  the  point  estimation  task  (fig.  \ref{fig:training_results_point}(a)).
  
To  compare  sampling-based  and  MAP  probabilistic  inferences,
we  evaluated  accuracies  in  terms  of  MSE  between  time-averaged  outputs  and  mean  value  of  the  posterior  distribution  given  various  inputs  as  follows:
  
\begin{equation}
	{\rm  MSE}  =  \frac{1}{N}\sum_{{\bf  x}}\bigl((\frac{1}{T}\sum_t  y(t))  -  \mu_{\rm  post}\bigr)^2
\end{equation}
  
Remarkably,  the  results  showed  the  RNN  is  more  accurate  in  the  sampling-based  task  than  in  the  MAP  task  (fig.\ref{fig:training_results_point}(c)).
In  other  words,  the  sampling-based  probabilistic  inference  on  the  posterior  distribution  works  as  an  inductive  bias  in  the  estimation  of  the  posterior  mean.  Notably,  this  difference  in  accuracy  between  MAP  estimation  and  sampling  is  subtle  in  the  FFNN  [Fig.  \ref{fig:training_results_point}(b)].
  
In  the  RNN  trained  to  perform  point  estimation,  each  pair  $(\mu_{\rm  post},  \sigma_{\rm  post})$  of  the  mean  and  standard  deviations  corresponds  to  a  fixed  point  ${\bf  h}^\ast(\mu_{\rm  post},  \sigma_{\rm  post})$  similar  to  the  sampling-based  task.
Since  it  is  implied  that  the  geometry  of  these  fixed  points  affects  the  accuracy  of  the  output  \citep{Cohen2020},  we  analyzed  it.
  
\begin{figure}[H]
	\centering
	\includegraphics[width=13.5cm]{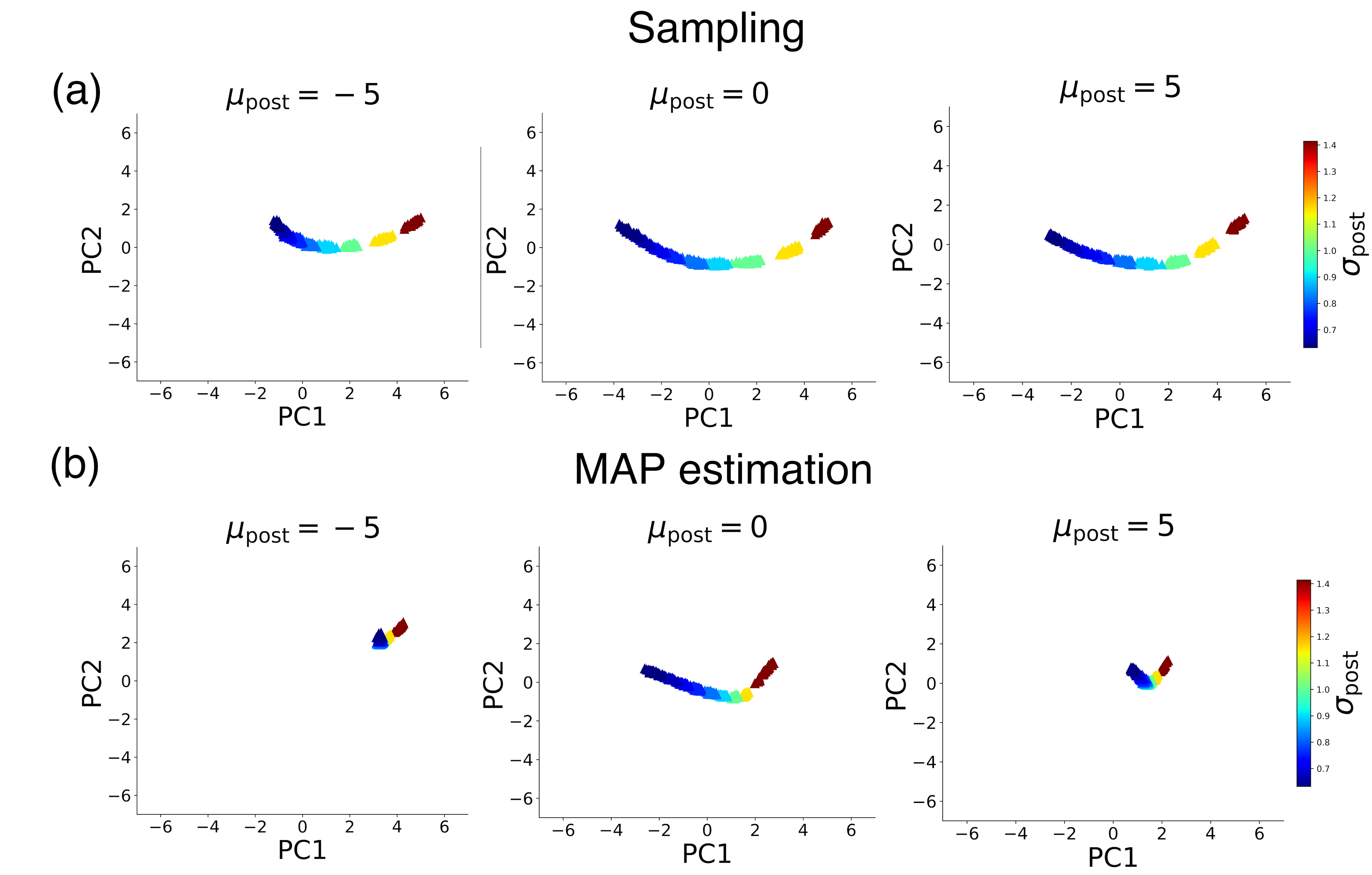}
	\caption{We  input  signals  under  fixed  posterior  mean  $\mu_{\rm  post}  =  0$,  computed  the  first  and  second  PCs  ${\bf  v}^0_1,{\bf  v}^0_2$  of  the  corresponding  fixed  points,  and  projected  fixed  points  ${\bf  h}^\ast(\mu_{\rm  post}  =  -5,  0,  5)$  given  input  signals  corresponding  to  $\mu_{\rm  post}  =  -5,  0,  5$  onto  ${\bf  v}^{\mu_{\rm  post}=0}_1,  {\bf  v}^{\mu_{\rm  post}  =  0}_2$.  (a)  Projected  fixed  points  in  the  RNN  trained  to  perform  sampling.  (b)  Projected  fixed  points  in  the  RNN  trained  to  perform  MAP  inference.  Although  the  relationships  between  $\sigma_{\rm  post}$  are  invariant  to  changes  in  $\mu_{\rm  post}$  in  the  former  NN,  they  are  variant  in  the  latter  NN.}
	\label{fig:inductive_bias}
\end{figure}
  
Fig.  \ref{fig:inductive_bias}  shows  that  ${\bf  h}^\ast(\mu_{\rm  post},  \sigma_{\rm  post})$  takes  different  positions  when  $\sigma_{\rm  post}$  varies,  even  if  $\mu_{\rm  post}$  is  constant.
By  contrast,  the  output  $y(t)$  is  required  to  be  fixed  when  $\mu_{\rm  post}$  is  fixed,  regardless  of  the  changes  in  $\sigma_{\rm  post}$.
To  fulfill  this  requirement,  the  NNs  need  readout  vector  ${\bf  W}_{\rm  out}$  to  be  aligned  orthogonal  to  the  direction  in  which  fixed  points  ${\bf  h}^\ast$  move  depending  on  $\sigma_{\rm  post}$.
Since  this  condition  should  be  satisfied  for  every  $\mu_{\rm  post}$,
changes  in  ${\bf  h}^\ast$  are  expected  to  be  small
when  $\mu_{\rm  post}$  is  fixed  and  $\sigma_{\rm  post}$  is  variable.
  
Assuming  that  ${\bf  h}^\ast$  is  sufficiently  low-dimensional,  the  varying  direction  of  ${\bf  h}^\ast$  is  characterized  by  PCs.
Hence,  for  a  detailed  characterization,  we  projected  a  set  (manifold)  of  fixed  points  ${\bf  h}^*(\mu_{\rm  post}=-5,0,5)$  given  $\mu_{\rm  post}=-5,0,5$  onto  the  first  and  second  PCs  computed  under  $\mu_{\rm  post}=0$.
If  the  direction  along  which  the  fixed  points  move  for  different  $\mu_{\rm  post}$  does  not  change  when  $\sigma_{\rm  post}$  varies,  the  projection  of  the  manifold  must  also  look  invariant.
As  shown  in  Fig.\ref{fig:inductive_bias},  the  NN  trained  on  the  sampling-based  task  satisfies  these  conditions  while  the  geometry  of  the  fixed-point  manifold  in  the  point  estimation  NN  is  apparently  variant  to  $\sigma_{\rm  post}$.
  
\begin{figure}
	\centering
	\includegraphics[width=10.5cm]{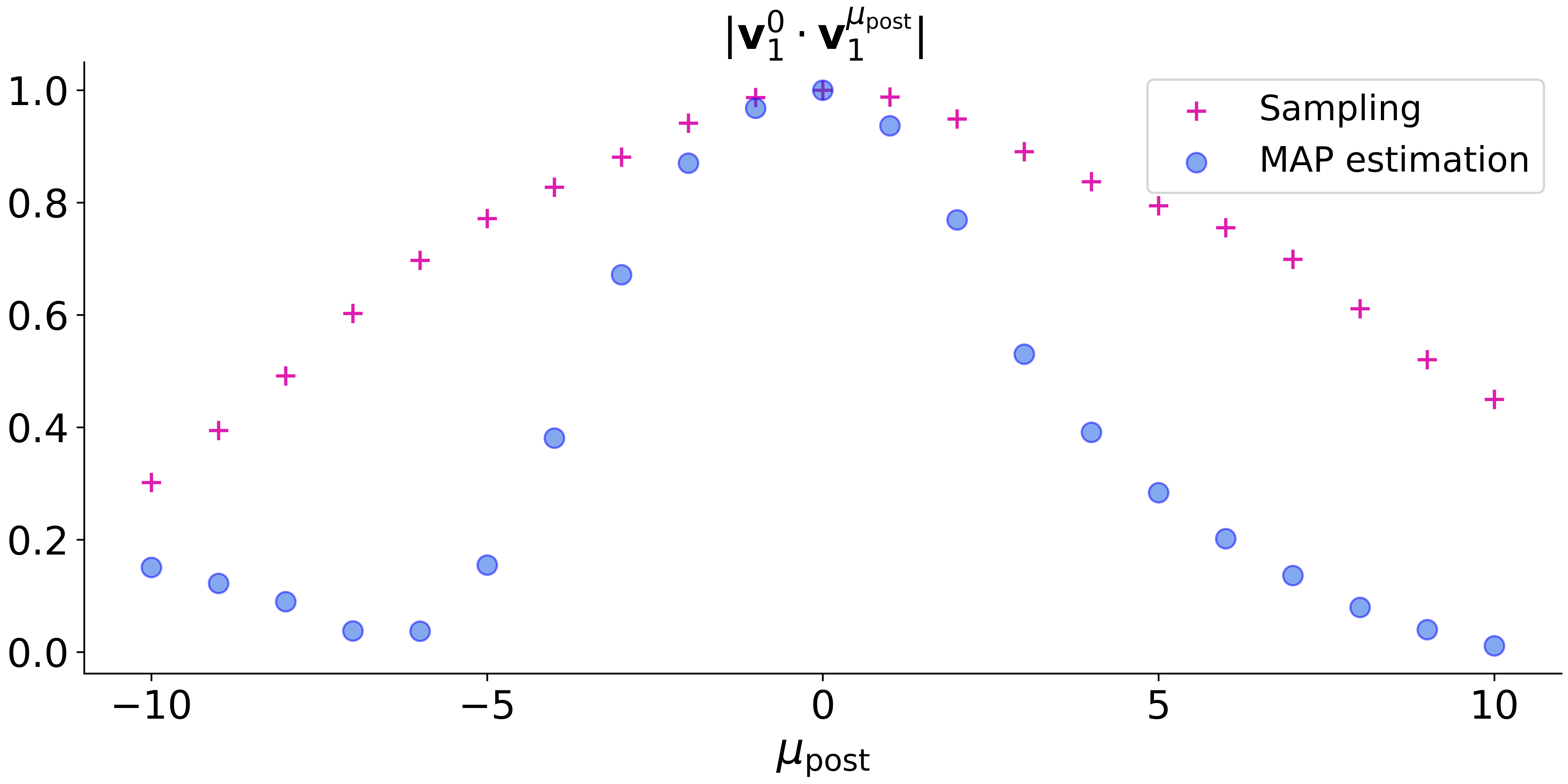}
	\caption{Absolute  inner  products  of  ${\bf  v}^0_1$  and  the  first  PCs  ${\bf  v}^{\mu_{\rm  post}}_1$  given  inputs  corresponding  to  different  $\mu_{\rm  post}$.
	The  bases  of  the  manifold  for  each  posterior  mean  changes  with  $\mu_{\rm  post}$.}
	\label{fig:inductive_bias2}
\end{figure}

Quantitatively,  these  geometric  characteristics  can  be  investigated  in  terms  of  absolute  inner  product  of  fixed  point  ${\bf  v}^{\mu_{\rm  post}}_1$  given  a  posterior  mean  ${\mu_{\rm  post}}$  and  ${\bf  v}^0_1$  as  $|{\bf  v}^0_1\cdot{\bf  v}^{\mu_{\rm  post}}_1|$.
$|{\bf  v}^0_1\cdot  {\bf  v}^{\mu_{\rm  post}}_1|$  takes  values  within  the  range  $(0,  1)$;  the  larger  this  value  is,  the  more  parallel  manifolds  for  different  posterior  means  are  aligned.
This  quantity  takes  larger  values  in  the  NN  trained  on  sampling-based  inference  (fig.  \ref{fig:inductive_bias2}),  which  means  that  the  manifolds  are  more  nearly  parallel  to  each  other  than  in  the  NN  for  point  estimation.
This  characteristic  implies  a  mechanism  wherein  the  NN  can  output  proper  values  depending  only  on  $\mu_{\rm  post}$,  independent  of  $\sigma_{\rm  post}$,  by  choosing  the  readout  vector  appropriately,  and  this  mechanism  underlies  the  superiority  of  the  RNN  trained  on  sampling  to  the  one  trained  on  MAP  estimation.
  
These  geometric  characteristics  in  sampling  might  be  due  to  the  learning  strategy  that  explicitly  includes  effects  of  $\sigma_{\rm  post}$  in  the  output  $y(t)$.
In  summary,  this  is  an  example  of  inductive  bias  \citep{Baxter,inductive_bias,dropout,HAUSSLER1988177}  in  which  a  constraint  leads  to  better  accuracy  in  estimating  $\mu_{\rm  post}$.
  
\section{Discussion}
In  this  study,  we  trained  two  types  of  ANNs,  the  FFNN  and  the  RNN,  on  the  cue-combination  task  to  investigate  their  dynamical  mechanisms  and  characteristics  of  performing  sampling-based  probabilistic  inferences.
To  achieve  this,  we  made  input  signals  encoding  external  cues  by  PPC  and  trained  the  ANNs  to  generate  the  output  as  a  sampling  of  posterior  probability  distribution  using  the  gradient  descent  scheme.
  
First,  the  results  showed  that  training  was  successful  for  both  ANNs.  Hence,  sampling-based  probabilistic  inferences  are  possible  using  the  ANNs.  However,  the  mechanisms  underlying  the  sampling  in  FFNN  and  RNN  differ:  the  FFNN  adjusts  amplitudes  of  the  fluctuation  of  the  input  signals  appropriately  when  they  are  feed-forwarded  into  the  hidden  layer  to  realize  a  sampling  from  the  posterior  distribution;  whereas,  in  the  RNN,  dynamics  in  the  hidden  layer,  which  can  be  decomposed  into  fixed  points  and  perturbations  around  them,  play  a  crucial  role  in  the  sampling.
Moreover,  it  was  revealed  that  the  RNN  employs  two  different  mechanisms  depending  on  the  posterior  variance.
When  the  posterior  variance  was  small,  the  angle  of  the  fluctuation  of  the  internal  state  ${\bf  h}(t)$  was  considered,  whereas  the  landscape  of  the  system  near  fixed  points  was  manipulated  when  the  posterior  variance  was  large.
It  is  also  suggested  that  this  difference  in  the  sampling  mechanisms  leads  to  the  superiority  of  the  RNN  in  terms  of  accuracy.
  
Further,  we  trained  the  FFNN  and  RNN  on  a  similar  cue-combination  task  but  in  a  MAP  estimation  setting,  not  sampling-based,  to  compare  the  sampling-based  probabilistic  inference  with  point  estimation.
The  results  showed  that,  in  the  RNN,  the  accuracy  of  estimating  mean  values  $\mu_{\rm  post}$  of  the  posterior  distribution  via  sampling  was  higher  than  that  via  point  estimation,  which  could  be  explained  by  differences  in  geometric  characteristics,  and  that  might  be  an  example  of  the  roles  of  the  sampling  as  an  inductive  bias.
The  changes  in  the  fixed-point  manifolds  constructed  by  varying  posterior  variance  $\sigma_{\rm  post}$  for  different  $\mu_{\rm  post}$  were  more  parallel  to  each  other  in  the  NN  trained  to  perform  the  sampling  than  in  that  trained  for  point  estimation,  leading  to  ease  for  the  readout  vector  alignment  to  be  invariant  to  changes  in  the  posterior  variance  $\sigma_{\rm  post}$,  which  accurizes  the  NN  \citep{PhysRevResearch.3.013176}.
This  characteristic  can  be  thought  to  be  obtained  under  constraints  in  which  the  NN  needs  to  constantly  behave  for  different  $\mu_{\rm  post}$.
  
In  this  study,  we  first  clarified  the  utility  of  sampling  in  information  processes.
In  addition,  it  is  suggested  that  the  sampling  can  contribute  to  good  performance  in  estimating  important  statistics,  such  as  the  MAP  inference,  not  only  in  the  estimation  of  ambiguities  of  information.
If  the  sampling  mechanisms  in  this  study  are  equipped  in  the  brain,  a  comparison  of  the  accuracy  between  point  estimation  and  sampling-based  inference  would  also  show  a  similar  result.
  
We  also  highlighted  differences  between  FFNNs  and  RNNs.
In  a  previous  study  \citep{Orhan2017},  FFNN  was  mainly  trained  on  the  point  estimation  setting,  and  differences  with  RNNs  were  not  discussed.
However,  considering  the  sampling,  there  were  clear  differences  between  them  in  terms  of  accuracy  and  mechanism.
These  results  can  provide  arguments  on  the  significance  of  recurrent  connections  among  neurons  in  the  brain  from  another  perspective.  Particularly,  this  study  suggested  important  results  to  investigate  what  roles  dynamical  characteristics  in  the  information  process  play  \citep{diverse_range}.
  
As  we  would  like  to  investigate  the  relationships  between  the  sampling  mechanism  and  PPC  for  the  first  step,  we  did  not  add  the  internal  noise  in  neural  dynamics.  However,  internal  noise  is  universally  present  in  the  brain,  and  it  probably  contributes  to  implementing  probabilistic  inference.  Indeed,  previous  studies  have  investigated  the  sampling  inference  of  RNN  driven  by  the  internal  noise\citep{Echeveste:2020aa,  10.1371/journal.pcbi.1005186}.  Hence,  how  internal  noise  and  noise  due  to  PPC  each  affect  sampling  needs  to  be  investigated  in  more  detail.  In  addition,  for  simplicity,  we  adopted  a  model  that  assumes  that  PPC  is  uncorrelated  between  input  neurons.  However,  since  a  large  body  of  work  shows  that  neurons  can  be  correlated  \citep{correlation_noise1,  correlation_noise2},  it  is  important  to  examine  PPC  that  correlates  among  input  neurons.  
  
The  model  we  used  for  our  research  is  an  idealized  model,  and  there  are  differences  in  the  way  the  brain  actually  works.
For  instance,  The  models  employed  in  this  study  were  trained  using  the  back-propagation  algorithm  \citep{bptt},  and  the  loss  function  was  KLD,  which  is  an  artificial  setting.
Although  it  is  still  a  matter  of  debate  as  to  what  kind  of  training  algorithm  is  equipped  in  the  actual  brain  \citep{backprop_brain1,  backprop_brain2},  many  reports  are  claiming  that  the  behaviors  of  ANNs  trained  using  the  back-propagation  algorithm  are  often  similar  to  activities  in  the  brain  \citep{deeplearning-neuroscience1,  deeplearning-neuroscience2,  deeplearning-neuroscience3},  even  though  it  is  said  to  be  physiologically  implausible  \citep{biologically_plausible1,  biologically_plausible2}.
Therefore,  we  expect  that  the  results  from  this  study  agree  with  the  dynamical  features  of  the  brain  known  from  previous  physiological  studies  or  those  that  will  be  revealed  in  future  studies.

\section*{Acknowledgments}
The  authors  would  like  to  express  their  gratitude  to  Kunihiko  Kaneko  and  Tetsuhiro  S.  Hatakeyama  for  their  comments  and  discussions.

\subsection*{Appendix  A:  Input  signals}
\begin{figure}[H]
	\centering
	\includegraphics[width=10.5cm]{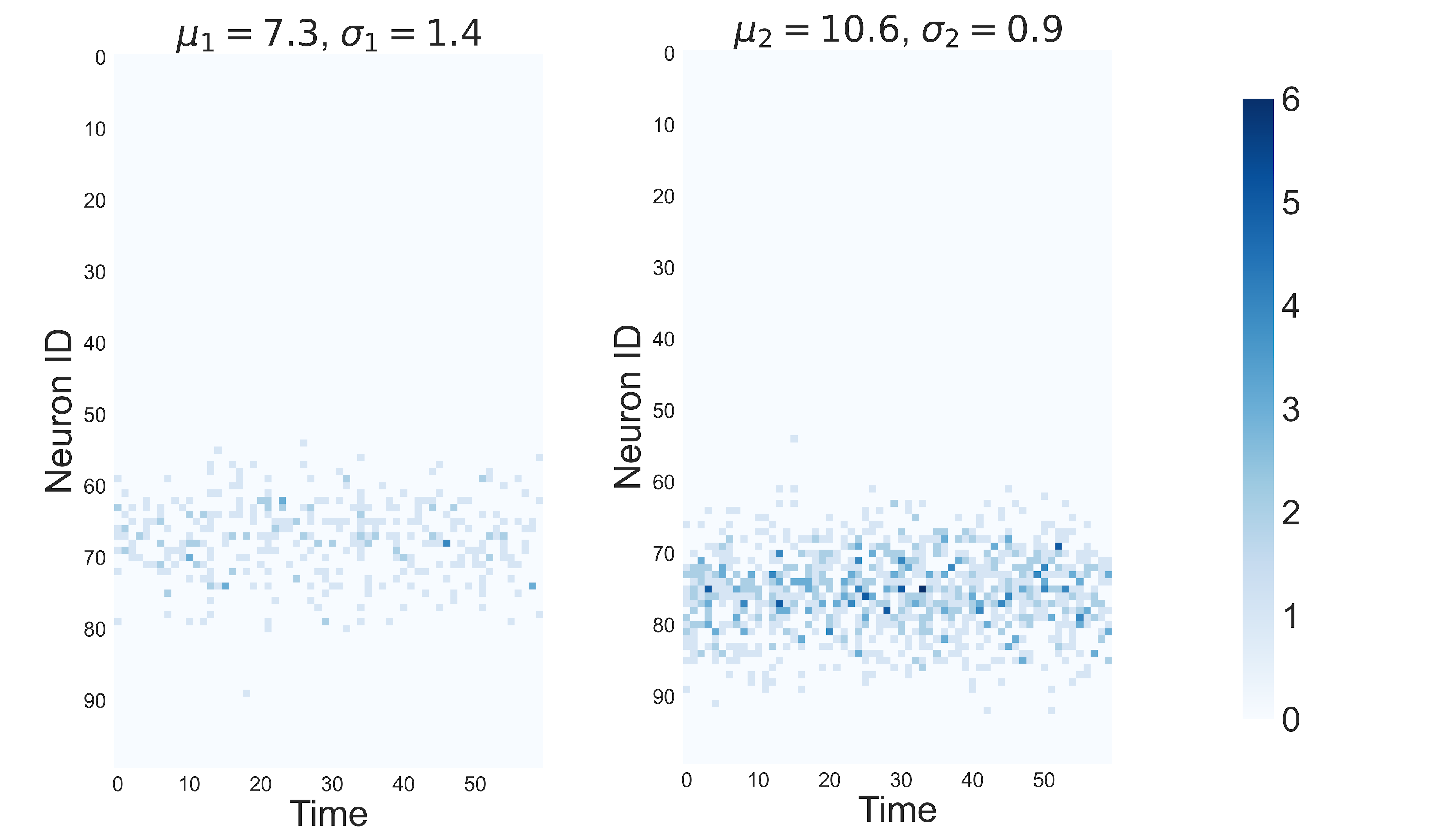}
	\caption{Example  of  an  input  signal.  Each  input  signals  encode  different  $\mu$,  $\sigma$.  They  are  generated  stochastically  according  to  Eqs.  (\ref{eq:poisson_variability1}),(\ref{eq:poisson_variability2}).}
	\label{fig:appendix1}
\end{figure}

The  input  signal  is  generated  stochastically  according  to  Eqs.  (\ref{eq:poisson_variability1}),  (\ref{eq:poisson_variability2}),  as  shown  in  Fig.\ref{fig:appendix1}.  The  two  signals  ${\bf  x}_1$  and  ${\bf  x}_2$  shown  in  the  figure  encode  the  input  information  with  different  mean  $\mu$  and  standard  deviation  $\sigma$,  respectively.  The  signal  of  ${\bf  x}_2$  has  a  smaller  $\sigma_2$  and  greater  certainty  of  information.  ${\bf  x}_2$  fires  more  strongly  than  ${\bf  x}_1$  because  $g_2$  is  larger  than  $g_1$.

\subsection*{Appendix  B:  Coordinate  Transformation  Task}
To  verify  that  sampling-based  inference  is  possible  for  other  tasks,  we  trained  the  coordinate  transformation  task  introduced  in  \citep{Beck15310}.  This  task  considers  the  same  setting  as  the  cue-combination  task,  where  signals  following  two  gaussian  distributions  are  input,  but  the  posterior  distribution  to  be  inferred  is  given  by  a  gaussian  distribution  $p_{\rm  post}  =  \mathcal{N}(\mu_{\rm  post},  \sigma^2_{\rm  post})$  with  the  following  mean  and  variance:
  
\begin{align}
	\label{eq:mu_post_appendix}
	\mu_{\rm  post}  &=  2\times\biggl(  \frac{\sigma_A^2}{\sigma_A^2  +  \sigma_B^2}\mu_B  +  \frac{\sigma_B^2}{\sigma_A^2  +  \sigma_B^2}\mu_A  \biggr)  \\
	\label{eq:sigma_post_appendix}
	\sigma_{\rm  post}^2  &=  \sigma_A^2  +  \sigma_B^2
\end{align}

\begin{figure}[H]
	\centering
	\includegraphics[width=13.5cm]{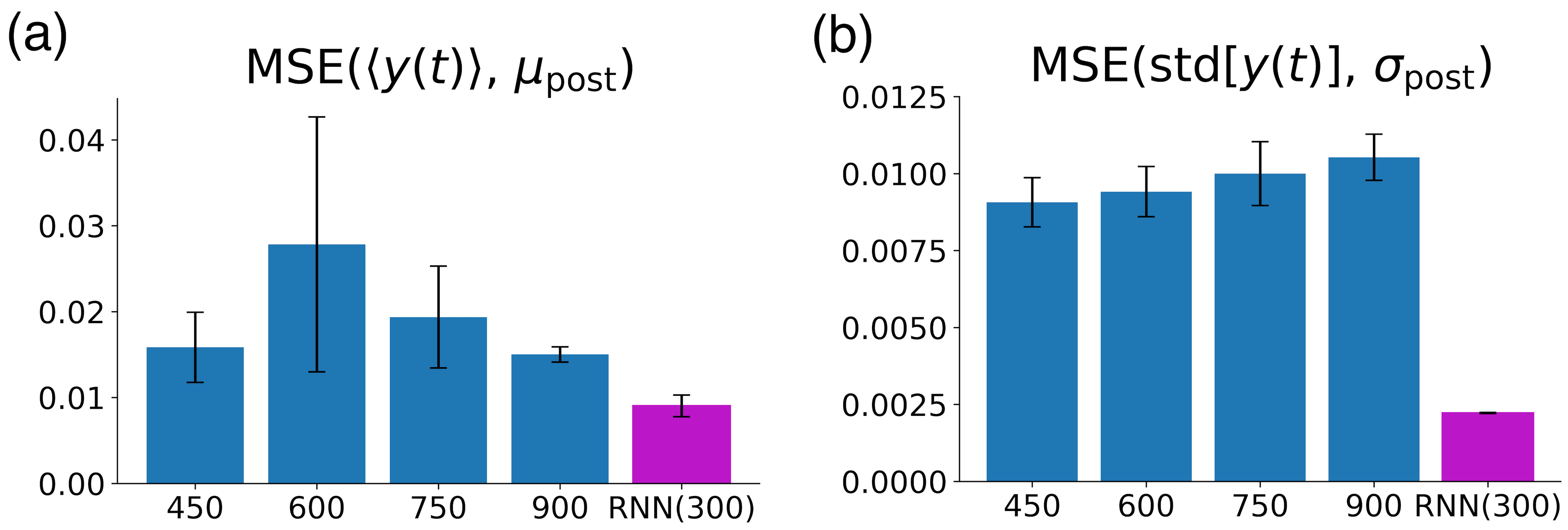}
	\caption{Comparison  of  accuracy  between  the  trained  RNN  and  the  trained  FFNN  with  different  network  sizes.  (a)  MSE  between  the  mean  $\mu_{\rm  post}$  of  the  posterior  distribution  and  the  mean  value  $\langle  y(t)  \rangle$  of  the  output.  (b)  MSE  of  the  standard  deviations  $\sigma_{\rm  post}$  of  the  posterior  distribution  and  the  standard  deviation  ${\rm  std}[y(t)]$  of  the  network.  Both  in  (a)  and  (b),  blue  bars  show  the  results  of  FFNN  and  purple  bar  shows  the  result  of  RNN.}
	\label{fig:appendix3}
\end{figure}

When  this  task  was  trained,  the  distribution  of  the  output  became  closer  to  the  posterior  distribution  through  learning(Fig.\ref{fig:appendix3}(a)),  and  the  mean  and  the  standard  deviation  of  the  output  had  values  close  to  the  mean  and  the  standard  deviation  of  the  posterior  distribution(Fig.\ref{fig:appendix3}(b)).  It  was  also  confirmed  that  the  output  $y(t)$  behaved  stochastically,  as  shown  in  Fig.\ref{fig:appendix3}(c).  From  these  results,  it  was  found  that  NNs  could  learn  sampling-based  inference  in  the  coordinate  transformation  task  setting  as  well  as  the  cue-combination  task.
  
\begin{figure}[H]
	\centering
	\includegraphics[width=13.5cm]{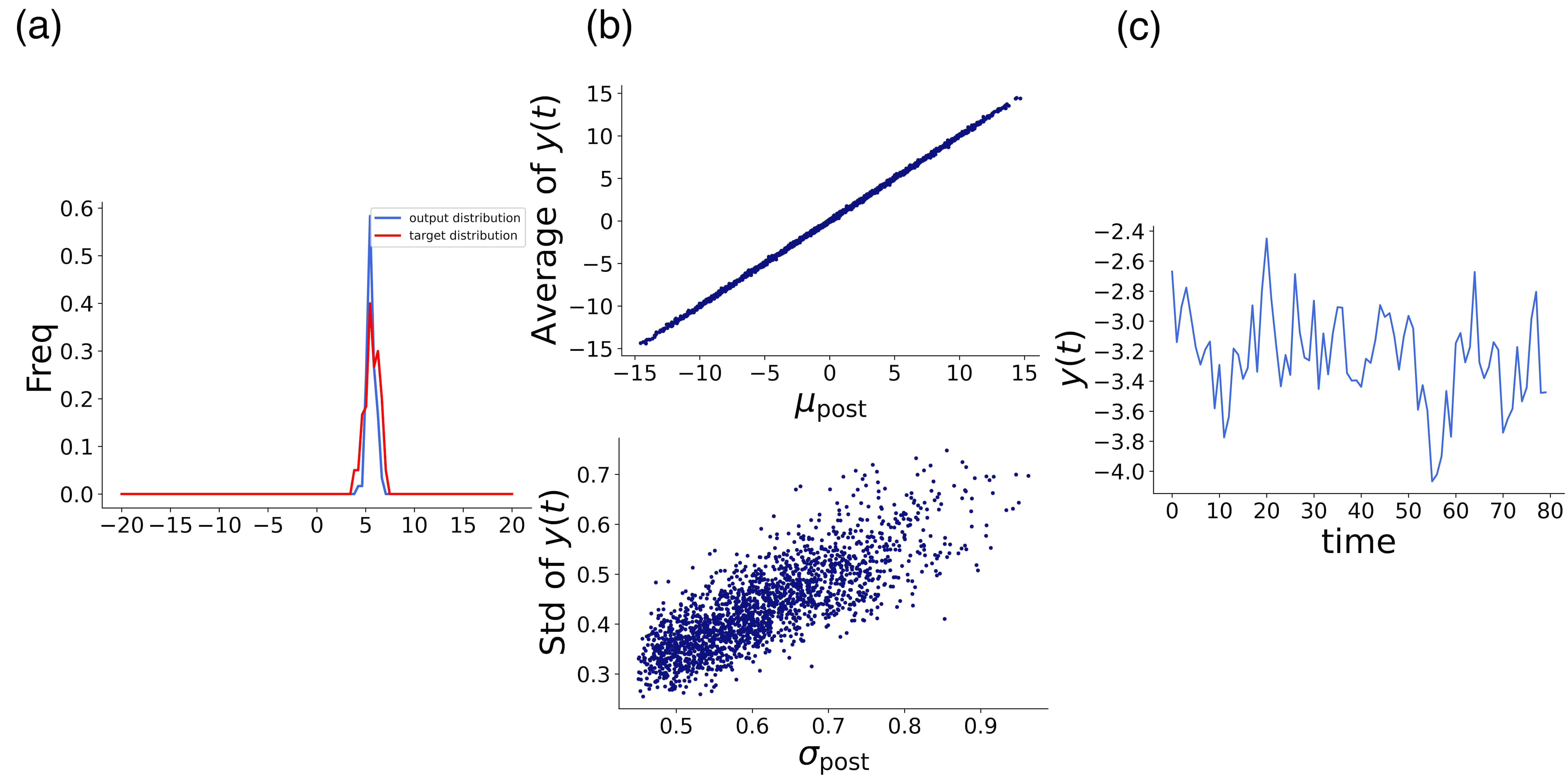}
	\caption{Behavior  of  the  trained  RNN  with  coordinate  transformation  task.  (a)Comparison  between  the  posterior  distribution  and  output  $y(t)$  of  the  RNN  after  training.  (b)(top)  Comparison  between  the  mean  of  the  posterior  distribution  $\mu_{\rm  post}$  and  the  mean  of  the  output  $y(t)$  of  the  network.  (bottom)  Comparison  between  the  std  of  the  posterior  distribution  $\sigma_{\rm  post}$  and  the  std  of  the  output  $y(t)$  of  the  network.  Each  scatter  corresponds  to  a  pair  of  input  signals  $({\bf  x}^A,{\bf  x}^B)$.  Here,  $\mu_A$  and  $\mu_B$  satisfies  $-15\leq  \mu_A,\mu_B  \leq  15$.  (c)  The  output  $y(t)$  of  the  network  given  a  pair  of  input  signals.  One  can  see  that  totally  stochastic  output,  neither  regular  nor  cyclic,  appears.}
	\label{fig:appendix2}
\end{figure}
  
\subsection*{Appendix  C:  Information  efficiency  in  the  NNs}
To  see  how  efficient  the  information  processing  performed  by  the  networks  is,  we  computed  information  losses  of  the  FFNN  and  RNN.
Information  loss  is  computed  as  KLD  from  the  optimal  posterior  distribution  to  the  histogram  of  outputs  of  a  network,  normalized  by  mutual  information  between  input  cue  and  label  data  \citep{Qamar20332}:
  
\begin{align*}
	\frac{D_{\rm  KL}[p(s|c^A,  c^B)\|q]}{{\rm  MI}(c^A,  c^B;  s)}  &=  \frac{\langle  \log  p(s|c^A,  c^B)  -  \log  q  \rangle_{{\rm  trial}}}{\langle  \log  p(s|c^A,  c^B)  -  \log  p(s)  \rangle_{{\rm  trial}}}.
\end{align*}
  
The  results  show  that  in  both  of  the  FFNN  and  the  RNN,  the  computed  information  losses  are  low,  which  suggest  these  networks  can  perform  efficient  information  processing.
By  comparing  the  information  loss  between  the  two  networks,  one  can  see  that  it  is  lower  in  the  RNN.
This  comparison  can  also  be  a  support  for  the  claim  that  the  RNN  performs  more  efficient  sampling-based  probabilistic  inference.
  
\begin{figure}[H]
	\centering
	\includegraphics[width=7cm]{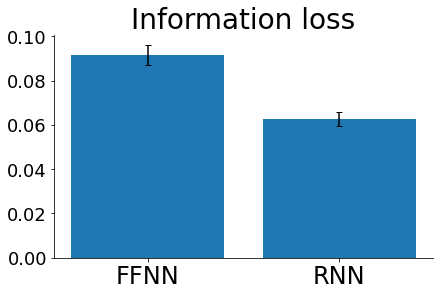}
	\caption{Information  losses  of  the  two  networks.  One  can  see  that  both  the  networks  can  perform  efficient  information  processing.  Especially,  the  information  loss  of  the  RNN  is  slightly  lower  than  the  FFNN.  The  results  suggest  that  the  RNN  conducts  more  efficient  information  processing  compared  to  the  FFNN,  from  the  perspective  of  information  theory.  $p$-value  calculated  by  statistical  test  was  $3  \times  10^{-20}$.}
	\label{fig:my_label}
\end{figure}
  
\subsection*{Appendix  D:  Effect  of  the  size  of  NNs}
To  confirm  that  the  difference  in  accuracy  between  FFNN  and  RNN  is  fundamental  one,  we  trained  the  cue-combination  task  in  FFNN  with  different  size  of  NNs  and  compared  the  accuracy.  The  FFNN  with  750  internal  neurons  and  the  RNN  with  300  internal  neurons  have  similar  number  of  parameters.
  
Although  the  accuracies  are  different  depending  on  the  size,  the  difference  between  FFNN  and  RNN,  especially  the  accuracy  on  standard  deviation,  is  remarkable(Fig.\ref{fig:appendix2}).  The  result  shows  that  there  is  a  fundamental  difference  in  accuracy  between  FFNN  and  RNN  as  discussed  in  the  main  part.
  
\bibliographystyle{apa}
  
\bibliography{ref}

\begin{thebibliography}{}

\bibitem[\protect\astroncite{Aitchison and
  Lengyel}{2016}]{10.1371/journal.pcbi.1005186}
Aitchison, L. and Lengyel, M. (2016).
\newblock The hamiltonian brain: Efficient probabilistic inference with
  excitatory-inhibitory neural circuit dynamics.
\newblock {\em PLOS Computational Biology}, 12(12):1--24.

\bibitem[\protect\astroncite{Angelaki et~al.}{2009}]{ANGELAKI2009452}
Angelaki, D.~E., Gu, Y., and DeAngelis, G.~C. (2009).
\newblock Multisensory integration: psychophysics, neurophysiology, and
  computation.
\newblock {\em Current Opinion in Neurobiology}, 19(4):452--458.
\newblock Sensory systems.

\bibitem[\protect\astroncite{Averbeck et~al.}{2006}]{correlation_noise1}
Averbeck, B.~B., Latham, P.~E., and Pouget, A. (2006).
\newblock Neural correlations, population coding and computation.
\newblock {\em Nature Reviews Neuroscience}, 7(5):358--366.

\bibitem[\protect\astroncite{B{\'a}nyai et~al.}{2019}]{Banyai2723}
B{\'a}nyai, M., Lazar, A., Klein, L., Klon-Lipok, J., Stippinger, M., Singer,
  W., and Orb{\'a}n, G. (2019).
\newblock Stimulus complexity shapes response correlations in primary visual
  cortex.
\newblock {\em PNAS}, 116(7):2723--2732.

\bibitem[\protect\astroncite{Barak}{2017}]{BARAK20171}
Barak, O. (2017).
\newblock Recurrent neural networks as versatile tools of neuroscience
  research.
\newblock {\em Current Opinion in Neurobiology}, 46:1--6.
\newblock Computational Neuroscience.

\bibitem[\protect\astroncite{Baxter}{2000}]{Baxter}
Baxter, J. (2000).
\newblock A model of inductive bias learning.
\newblock {\em J. Artif. Int. Res.}, 12(1):149–198.

\bibitem[\protect\astroncite{Beck et~al.}{2011}]{Beck15310}
Beck, J.~M., Latham, P.~E., and Pouget, A. (2011).
\newblock Marginalization in neural circuits with divisive normalization.
\newblock {\em Journal of Neuroscience}, 31(43):15310--15319.

\bibitem[\protect\astroncite{Beck et~al.}{2008}]{BECK20081142}
Beck, J.~M., Ma, W.~J., Kiani, R., Hanks, T., Churchland, A.~K., Roitman, J.,
  Shadlen, M.~N., Latham, P.~E., and Pouget, A. (2008).
\newblock Probabilistic population codes for bayesian decision making.
\newblock {\em Neuron}, 60(6):1142--1152.

\bibitem[\protect\astroncite{Bengio et~al.}{2015}]{biologically_plausible1}
Bengio, Y., Lee, D., Bornschein, J., and Lin, Z. (2015).
\newblock Towards biologically plausible deep learning.
\newblock {\em ArXiv}, abs/1502.04156.

\bibitem[\protect\astroncite{Chung et~al.}{2018}]{PhysRevX.8.031003}
Chung, S., Lee, D.~D., and Sompolinsky, H. (2018).
\newblock Classification and geometry of general perceptual manifolds.
\newblock {\em Phys. Rev. X}, 8:031003.

\bibitem[\protect\astroncite{Cohen et~al.}{2020}]{Cohen2020}
Cohen, U., Chung, S., Lee, D.~D., and Sompolinsky, H. (2020).
\newblock Separability and geometry of object manifolds in deep neural
  networks.
\newblock {\em Nature Communications}, 11(1):746.

\bibitem[\protect\astroncite{Doya et~al.}{2007}]{Doya2007-yy}
Doya, K., Ishii, S., Pouget, A., and Rao, R. P.~N. (2007).
\newblock {\em Bayesian Brain: Probabilistic Approaches to Neural Coding}.
\newblock MIT Press.

\bibitem[\protect\astroncite{Echeveste et~al.}{2020}]{Echeveste:2020aa}
Echeveste, R., Aitchison, L., Hennequin, G., and Lengyel, M. (2020).
\newblock Cortical-like dynamics in recurrent circuits optimized for
  sampling-based probabilistic inference.
\newblock {\em Nature Neuroscience}, 23(9):1138--1149.

\bibitem[\protect\astroncite{Ernst and Banks}{2002}]{Ernst2002}
Ernst, M.~O. and Banks, M.~S. (2002).
\newblock Humans integrate visual and haptic information in a statistically
  optimal fashion.
\newblock {\em Nature}, 415(6870):429--433.

\bibitem[\protect\astroncite{Haefner et~al.}{2016}]{HAEFNER2016649}
Haefner, R.~M., Berkes, P., and Fiser, J. (2016).
\newblock Perceptual decision-making as probabilistic inference by neural
  sampling.
\newblock {\em Neuron}, 90(3):649--660.

\bibitem[\protect\astroncite{Haussler}{1988}]{HAUSSLER1988177}
Haussler, D. (1988).
\newblock Quantifying inductive bias: Ai learning algorithms and valiant's
  learning framework.
\newblock {\em Artificial Intelligence}, 36(2):177--221.

\bibitem[\protect\astroncite{Helmbold and Long}{2015}]{dropout}
Helmbold, D.~P. and Long, P.~M. (2015).
\newblock On the inductive bias of dropout.
\newblock {\em Journal of Machine Learning Research}, 16(105):3403--3454.

\bibitem[\protect\astroncite{Jolliffe and Cadima}{2016}]{PCA}
Jolliffe, I.~T. and Cadima, J. (2016).
\newblock Principal component analysis: a review and recent developments.
\newblock {\em Philosophical Transactions of the Royal Society A: Mathematical,
  Physical and Engineering Sciences}, 374(2065):20150202.

\bibitem[\protect\astroncite{Kingma and Ba}{2014}]{adam}
Kingma, D.~P. and Ba, J. (2014).
\newblock Adam: A method for stochastic optimization.

\bibitem[\protect\astroncite{Knill and Pouget}{2004}]{KNILL2004712}
Knill, D.~C. and Pouget, A. (2004).
\newblock The bayesian brain: the role of uncertainty in neural coding and
  computation.
\newblock {\em Trends in Neurosciences}, 27(12):712--719.

\bibitem[\protect\astroncite{LeCun et~al.}{2015}]{DeepLearning}
LeCun, Y., Bengio, Y., and Hinton, G. (2015).
\newblock Deep learning.
\newblock {\em Nature}, 521:436--444.

\bibitem[\protect\astroncite{Lillicrap et~al.}{2016}]{biologically_plausible2}
Lillicrap, T.~P., Cownden, D., Tweed, D.~B., and Akerman, C.~J. (2016).
\newblock Random synaptic feedback weights support error backpropagation for
  deep learning.
\newblock {\em Nature Communications}, 7(1):13276.

\bibitem[\protect\astroncite{Lillicrap et~al.}{2020}]{backprop_brain1}
Lillicrap, T.~P., Santoro, A., Marris, L., Akerman, C.~J., and Hinton, G.
  (2020).
\newblock Backpropagation and the brain.
\newblock {\em Nature Reviews Neuroscience}, 21(6):335--346.

\bibitem[\protect\astroncite{Ma et~al.}{2006}]{Ma2006}
Ma, W.~J., Beck, J.~M., Latham, P.~E., and Pouget, A. (2006).
\newblock Bayesian inference with probabilistic population codes.
\newblock {\em Nature Neuroscience}, 9(11):1432--1438.

\bibitem[\protect\astroncite{Ma et~al.}{2008}]{MA2008217}
Ma, W.~J., Beck, J.~M., and Pouget, A. (2008).
\newblock Spiking networks for bayesian inference and choice.
\newblock {\em Current Opinion in Neurobiology}, 18(2):217--222.
\newblock Cognitive neuroscience.

\bibitem[\protect\astroncite{Maheswaranathan
  et~al.}{2019}]{NEURIPS2019_5f5d4720}
Maheswaranathan, N., Williams, A., Golub, M., Ganguli, S., and Sussillo, D.
  (2019).
\newblock Universality and individuality in neural dynamics across large
  populations of recurrent networks.
\newblock In Wallach, H., Larochelle, H., Beygelzimer, A., d\textquotesingle
  Alch\'{e}-Buc, F., Fox, E., and Garnett, R., editors, {\em Advances in Neural
  Information Processing Systems}, volume~32. Curran Associates, Inc.

\bibitem[\protect\astroncite{Mante et~al.}{2013}]{deeplearning-neuroscience3}
Mante, V., Sussillo, D., Shenoy, K.~V., and Newsome, W.~T. (2013).
\newblock Context-dependent computation by recurrent dynamics in prefrontal
  cortex.
\newblock {\em Nature}, 503(7474):78--84.

\bibitem[\protect\astroncite{Merfeld et~al.}{1999}]{Merfeld1999}
Merfeld, D.~M., Zupan, L., and Peterka, R.~J. (1999).
\newblock Humans use internal models to estimate gravity and linear
  acceleration.
\newblock {\em Nature}, 398(6728):615--618.

\bibitem[\protect\astroncite{Moreno-Bote et~al.}{2014}]{correlation_noise2}
Moreno-Bote, R., Beck, J., Kanitscheider, I., Pitkow, X., Latham, P., and
  Pouget, A. (2014).
\newblock Information-limiting correlations.
\newblock {\em Nature Neuroscience}, 17(10):1410--1417.

\bibitem[\protect\astroncite{Moreno-Bote et~al.}{2011}]{Moreno-Bote12491}
Moreno-Bote, R., Knill, D.~C., and Pouget, A. (2011).
\newblock Bayesian sampling in visual perception.
\newblock {\em Proceedings of the National Academy of Sciences},
  108(30):12491--12496.

\bibitem[\protect\astroncite{Nair and Hinton}{2010}]{ReLU}
Nair, V. and Hinton, G.~E. (2010).
\newblock Rectified linear units improve restricted boltzmann machines.
\newblock In Fürnkranz, J. and Joachims, T., editors, {\em ICML}, pages
  807--814. Omnipress.

\bibitem[\protect\astroncite{Orbán et~al.}{2016}]{ORBAN2016530}
Orbán, G., Berkes, P., Fiser, J., and Lengyel, M. (2016).
\newblock Neural variability and sampling-based probabilistic representations
  in the visual cortex.
\newblock {\em Neuron}, 92(2):530--543.

\bibitem[\protect\astroncite{Orhan and Ma}{2017}]{Orhan2017}
Orhan, A.~E. and Ma, W.~J. (2017).
\newblock Efficient probabilistic inference in generic neural networks trained
  with non-probabilistic feedback.
\newblock {\em Nature Communications}, 8(1):138.

\bibitem[\protect\astroncite{Orhan and Ma}{2019}]{diverse_range}
Orhan, A.~E. and Ma, W.~J. (2019).
\newblock A diverse range of factors affect the nature of neural
  representations underlying short-term memory.
\newblock {\em Nature Neuroscience}, 22(2):275--283.

\bibitem[\protect\astroncite{Qamar et~al.}{2013}]{Qamar20332}
Qamar, A.~T., Cotton, R.~J., George, R.~G., Beck, J.~M., Prezhdo, E., Laudano,
  A., Tolias, A.~S., and Ma, W.~J. (2013).
\newblock Trial-to-trial, uncertainty-based adjustment of decision boundaries
  in visual categorization.
\newblock {\em Proceedings of the National Academy of Sciences},
  110(50):20332--20337.

\bibitem[\protect\astroncite{Richards
  et~al.}{2019}]{deeplearning-neuroscience1}
Richards, B.~A., Lillicrap, T.~P., Beaudoin, P., Bengio, Y., Bogacz, R.,
  Christensen, A., Clopath, C., Costa, R.~P., de~Berker, A., Ganguli, S.,
  Gillon, C.~J., Hafner, D., Kepecs, A., Kriegeskorte, N., Latham, P., Lindsay,
  G.~W., Miller, K.~D., Naud, R., Pack, C.~C., Poirazi, P., Roelfsema, P.,
  Sacramento, J., Saxe, A., Scellier, B., Schapiro, A.~C., Senn, W., Wayne, G.,
  Yamins, D., Zenke, F., Zylberberg, J., Therien, D., and Kording, K.~P.
  (2019).
\newblock A deep learning framework for neuroscience.
\newblock {\em Nature Neuroscience}, 22(11):1761--1770.

\bibitem[\protect\astroncite{Susman et~al.}{2021}]{PhysRevResearch.3.013176}
Susman, L., Mastrogiuseppe, F., Brenner, N., and Barak, O. (2021).
\newblock Quality of internal representation shapes learning performance in
  feedback neural networks.
\newblock {\em Phys. Rev. Research}, 3:013176.

\bibitem[\protect\astroncite{Sussillo and Barak}{2013}]{fixed_point}
Sussillo, D. and Barak, O. (2013).
\newblock {Opening the Black Box: Low-Dimensional Dynamics in High-Dimensional
  Recurrent Neural Networks}.
\newblock {\em Neural Computation}, 25(3):626--649.

\bibitem[\protect\astroncite{Svozil et~al.}{1997}]{SVOZIL199743}
Svozil, D., Kvasnicka, V., and Pospichal, J. (1997).
\newblock Introduction to multi-layer feed-forward neural networks.
\newblock {\em Chemometrics and Intelligent Laboratory Systems}, 39(1):43--62.

\bibitem[\protect\astroncite{Tanabe}{2013}]{Tanabe:2013aa}
Tanabe, S. (2013).
\newblock Population codes in the visual cortex.
\newblock {\em Neuroscience research}, 76(3):101--105.

\bibitem[\protect\astroncite{Tolhurst et~al.}{1983}]{TOLHURST1983775}
Tolhurst, D., Movshon, J., and Dean, A. (1983).
\newblock The statistical reliability of signals in single neurons in cat and
  monkey visual cortex.
\newblock {\em Vision Research}, 23(8):775--785.

\bibitem[\protect\astroncite{van Beers et~al.}{1999}]{cue_combination}
van Beers, R.~J., Sittig, A.~C., and Gon, J. J. D. v.~d. (1999).
\newblock Integration of proprioceptive and visual position-information: An
  experimentally supported model.
\newblock {\em Journal of Neurophysiology}, 81(3):1355--1364.
\newblock PMID: 10085361.

\bibitem[\protect\astroncite{Vyas et~al.}{2020}]{Vyas2020-fp}
Vyas, S., Golub, M.~D., Sussillo, D., and Shenoy, K.~V. (2020).
\newblock Computation through neural population dynamics.
\newblock {\em Annu. Rev. Neurosci.}, 43:249--275.

\bibitem[\protect\astroncite{Walker et~al.}{2020}]{Walker:2020aa}
Walker, E.~Y., Cotton, R.~J., Ma, W.~J., and Tolias, A.~S. (2020).
\newblock A neural basis of probabilistic computation in visual cortex.
\newblock {\em Nature Neuroscience}, 23(1):122--129.

\bibitem[\protect\astroncite{{Werbos}}{1990}]{bptt}
{Werbos}, P.~J. (1990).
\newblock Backpropagation through time: what it does and how to do it.
\newblock {\em Proceedings of the IEEE}, 78(10):1550--1560.

\bibitem[\protect\astroncite{Whigham}{1995}]{inductive_bias}
Whigham, P. (1995).
\newblock Inductive bias and genetic programming.
\newblock {\em IET Conference Proceedings}, pages 461--466(5).

\bibitem[\protect\astroncite{Whittington and Bogacz}{2019}]{backprop_brain2}
Whittington, J.~C. and Bogacz, R. (2019).
\newblock Theories of error back-propagation in the brain.
\newblock {\em Trends in Cognitive Sciences}, 23(3):235--250.

\bibitem[\protect\astroncite{Yang and Wang}{2020}]{deeplearning-neuroscience2}
Yang, G.~R. and Wang, X.-J. (2020).
\newblock Artificial neural networks for neuroscientists: A primer.
\newblock {\em Neuron}, 107(6):1048--1070.

\end{thebibliography}

\end{document}